\documentclass[12pt]{article} \topmargin-28mm \textheight245mm \textwidth180mm \oddsidemargin-8mm

\errorstopmode

\usepackage{graphicx}
\usepackage{amsmath}
\usepackage{amssymb}

\newcommand{\dd}{{\rm d}}
\newcommand{\ii}{{\rm i}}
\newcommand{\fii}{\varphi}
\newcommand{\rhho}{N}
\newcommand{\ellzero}{L_{0}}  
\newcommand{\kay}{L}      
\newcommand{\arcosh}{{\rm arcosh\,}}
\newcommand{\e}{{\rm e}}

\begin{document}

\title{Absolute instruments and perfect imaging \\in geometrical optics}

\author{Tom\'a\v{s} Tyc, Lenka Herz\'anov\'{a},\\ Martin \v
  Sarbort  and Klaus Bering\\
Institute of Theoretical Physics and Astrophysics,\\
Masaryk University, Kotl\'a\v rsk\'a 2, 61137 
Brno, Czech Republic}

\date{28 June 2011}

\maketitle

\begin{abstract}

We investigate imaging by spherically symmetric absolute instruments that
provide perfect imaging in the sense of geometrical optics.  We derive a number
of properties of such devices, present a general method for designing them and
use this method to propose several new absolute instruments, in particular a
lens providing a stigmatic image of an optically homogeneous region and having
a moderate refractive index range.


\end{abstract}
\maketitle

\section{Introduction}

In the last decade, perfect imaging has become one of the hot topics in
optics. This was triggered by a discovery of Sir J. Pendry~\cite{Pendry2000}
who showed that a slab of material with negative refractive index can focus
light to a spot much smaller than the wavelength of light.  Modern
metamaterials with carefully designed electric and magnetic
responses~\cite{Soukoulis2007} provided a suitable experimental background for
testing such super-resolution~\cite{Fang2005-neg_n-silver_imaging}, but it has
turned out that there are severe limitations due to a strong absorption in
negatively refracting materials~\cite{Stockman2007}.  This started a search for
other devices that could provide super-resolution but would not suffer from the
disadvantages related to negative refraction.  It has turned out recently that
such devices do exist, which was demonstrated both theoretically
\cite{Ulf2009-fisheye,Ulf2010-fisheye} and experimentally \cite{Ma2011} for a
well-known optical device, Maxwell's fish eye \cite{MFE}. Optical imaging with
super-resolution is generally called perfect imaging and the corresponding
devices are called perfect lenses.

The concept of perfect imaging has become important also in geometrical optics
where limitations of optical instruments are of a different kind. There, the
diffraction limit of resolution is not the subject of investigation, but
instead one seeks to minimise or even eliminate the optical aberrations of the
device.  Remarkably, there exist devices in which this elimination can be made
perfect; they are called absolute instruments~\cite{BornWolf}. An important
subset of absolute instruments is formed by devices that produce images
geometrically similar to the imaged objects. Within geometrical optics, such
images are called perfect~\cite{BornWolf}. Hence the meaning of ``perfect
imaging'' is different in geometrical and in wave optics. In particular, an
imaging device that is perfect from the point of view of geometrical optics may
or may not be perfect from the point of view of wave optics, and vice
versa. The only devices that are known to produce images perfect in both senses
are Pendry's slab~\cite{Pendry2000} and Maxwell's fisheye surrounded by a
mirror~\cite{Ulf2009-fisheye,Ma2011}; it remains unknown whether other lenses
producing stigmatic images also provide super-resolution. However, as the
answer is very likely to be positive, a search for new absolute instruments
with interesting and useful properties is highly desirable because such devices
might provide unprecedented resolution and find applications in imaging or
lithography.

In this paper, we focus on absolute instruments and perfect imaging from the
point of view of geometrical optics.  We analyse general properties of imaging
by spherically symmetric absolute instruments and develop a method for
designing such devices. We then use this method for proposing several devices
with interesting properties.  

The paper is organised as follows. In Sec.~\ref{theorem} we analyse spherically
symmetric absolute instruments and derive several general results. In
Sec.~\ref{inversion} we solve the inverse scattering problem for spatially
confined rays and based on this result we develop a method for designing
absolute instruments in Sec.~\ref{method}.  In Sec.~\ref{examples} we give some
simple examples of absolute instruments and in Secs.~\ref{homogeneous}
and~\ref{multiple} we discuss devices with homogeneous regions and multiple
image points, respectively. We conclude in Sec.~\ref{conclusion}.

\section{Properties of spherically symmetric absolute instruments}
\label{theorem}

In this paper we will consider isotropic spherically symmetric refractive index
profiles $n(r)$. Such situations have a great advantage: if some point at
radius $r$ is imaged stigmatically, then the same is true for all points at the
same radius.  At the same time, usually the object to be imaged must be
embedded directly into the optical medium and the rays emerge in all directions
from the object and come from all directions to the image.  This is quite
different from the usual imaging e.g. by a camera or a telescope where rays
propagate more or less in one direction.

\subsection{Angular momentum and turning parameter}

It is well known (and follows for example from the analogy between geometrical
optics and classical mechanics) that in spherically symmetrical refractive
index profiles $n(r)$ light rays propagate in a plane containing the centre of
symmetry. This is a consequence of conservation of the quantity analogous to
mechanical angular momentum, the magnitude of which is~\cite{BornWolf}
\begin{equation}
      L=rn(r)\sin\alpha\,,
\label{L}\end{equation}
where $\alpha$ is the angle between the tangent to the particle trajectory and
the radius vector. Motion in spherically symmetric index distributions $n(r)$ is
therefore effectively two-dimensional, and we will treat it as such unless
otherwise stated.

For a spherically symmetric medium, an important role is played by the radially
normalised index function
\begin{equation}
\rhho(r) = rn(r)\,,
\label{rhhodef}\end{equation}
which is also known as the turning parameter~\cite{Hendi2006,Ulf-Thomas-book}.
Consider a light ray propagating with angular momentum $L$.
It follows from Eqs.~(\ref{L}) and~(\ref{rhhodef}) that
$L\!=\!\rhho(r)\sin\alpha$ and hence
the ray can propagate only in the regions where $L\leq\rhho(r)$. When it gets
to the point where $L\!=\!\rhho(r)$, then $\alpha=\pi/2$ and the ray propagates
purely in the angular direction. Such a point is called a turning point.

\subsection{Stigmatic images and absolute instruments}

According to Principles of Optics by M. Born and E.  Wolf~\cite{BornWolf},
an absolute instrument is a device that images stigmatically a three-dimensional
domain of space. A stigmatic image of a point A is a point B through which an
infinity of rays emerging from A pass.

We shall distinguish between two cases of a stigmatic image. In the first case,
the image of a point A is formed at B by all rays emerging from A into some
nonzero solid angle. We shall then call B a {\em strong stigmatic image} (or
simply {\em strong image}) of A.  In the second case, although there is an
infinite number of rays going from A to B, these rays constitute just a zero
solid angle. Then we shall call B a {\em weak [stigmatic] image}. For example, a
cylindrical lens can form a whole line of weak images of a given point but no
strong image.

\subsection{Mutual position of an object and its image}

Suppose we have an absolute instrument with refractive index $n(r)$ that
stigmatically images a point A with radial coordinate $r_{\rm A}$ to a
point B with radial coordinate $r_{\rm B}$. This means that an infinite
number of rays emerging from A meet at B~\cite{BornWolf}. Imagine we shift the
point A infinitesimally in the angular direction by an angle $\dd\fii$ to a new
position A$'$ separated from A by distance $r_{\rm A}\dd\fii$. From the
spherical symmetry it follows that the image B gets shifted by the same angle
$\dd\fii$ to a new position B$'$ separated from B by $r_{\rm B}\dd\fii$. Now
Maxwell's theorem for absolute instruments~\cite{BornWolf} states that the
optical length of any curve and of its image is the same, which we can apply to
the lines AA$'$ and BB$'$. Cancelling the angle $\dd\fii$, we obtain
\begin{equation}
  n(r_{\rm A})\,  r_{\rm A}=  n(r_{\rm B})\,  r_{\rm B} \,.
\label{angular}\end{equation}
Assuming that we shift the point A in the radial direction by $|\dd r_{\rm A}|$
to a point A$''$ instead, the point B is shifted also in the radial direction
to the point B$''$. This follows from the fact that imaging by absolute
instruments is conformal~\cite{BornWolf}, another consequence of Maxwell's
theorem: if the angle A$'$AA$''$ is $\pi/2$, so must be the angle B$'$BB$''$.
From Maxwell's theorem applied to the lines AA$''$ and BB$''$ we then get
\begin{equation}
  n(r_{\rm A})|\dd r_{\rm A}|= n(r_{\rm B})|\dd r_{\rm B}|\,.
\label{radial}\end{equation}
Dividing Eq.~(\ref{radial}) by Eq.~(\ref{angular}), we get 
\begin{equation}
  \frac{\dd r_{\rm A}}{r_{\rm A}}=\pm \frac{\dd r_{\rm B}}{r_{\rm B}}
\label{plusminus}\end{equation}
 with the solutions
\begin{eqnarray}
  r_{\rm B}&=&kr_{\rm A} \label{first}\\
  r_{\rm B}&=&\frac k{r_{\rm A}}\,
\label{second}\end{eqnarray}
corresponding to plus and minus sign in Eq.~(\ref{plusminus}), respectively,
and an integration constant $k>0$.

Another question is related to the mutual angular position of an object and its
image. Can there be a situation that the object A, its strong stigmatic image B
and the origin O (centre of symmetry of the lens) do not lie on a single
straight line?  Imagine such a situation. Since B is a strong image of A, a
full solid angle of rays starting from A pass through B. We can rotate these
rays around the line OA by some angle $\fii$, which also moves the point B to a
new position B$'$. The rotated rays now pass through B$'$ which, due to the
spherical symmetry of the lens, must also be a strong stigmatic image of
A. This would mean that A has strong images along the whole circle that is
obtained by rotation of the point B around the axis OA, which is clearly
impossible.  So we must conclude that the points A, B and O lie on a single
straight line. This way the image B is either on the exactly same side from O
as is A or on the exactly opposite side.

Note that this argument is not valid in two dimensions where the mutual
position of a point and its image in a rotationally-symmetric absolute
instrument is less restricted. For the same reason this argument is not valid
in 3D for weak images.

\subsection{Value of the constant $k$ in Eq.~(\ref{first})}

An interesting question is related to the possible values of the constant $k$
in Eq.~(\ref{first}). We conjecture that the only possibility is
$k=1$. Although we have not been able to show this in the completely general
situation, the proof can be given in two practically important cases. For the
first case, assume that the constant $k$ is the same for all the points that
are imaged by the lens.  But if B is an image of A, then also A is an image of
B, so $r_{\rm A}=kr_{\rm B}$ must also hold along with Eq.~(\ref{first}), from
which then $k=1$ follows.

The assumption of the second case is that among the rays that contribute to the
image B of a point A there is also the ray for which A is a turning point. In
other words, we assume that a ray starting from the point A in the angular
direction makes it to the image of A formed at the point B. Since
Eq.~(\ref{angular}) can be rewritten as $N(r_{\rm A})=N(r_{\rm B})$, it follows
that the point B is a turning point for the same ray as well.  Next we use
Eq.~(\ref{plusminus}) with the plus sign from which Eq.~(\ref{first}) has been
derived, i.e., $\dd r_{\rm A}/r_{\rm A}=\dd r_{\rm B}/r_{\rm B}$. From this
equation combined with $N(r_{\rm A})=N(r_{\rm B})$ then follows
\begin{equation}
  r_{\rm A}\, \frac{\dd N(r_{\rm A})}{\dd r_{\rm A}}=
  r_{\rm B}\, \frac{\dd N(r_{\rm B})}{\dd r_{\rm B}}\,.
\label{dN_dr}\end{equation}
We see that the derivative $\dd N/\dd r$ has the same sign at both $r_{\rm A}$
and $r_{\rm B}$. However, if $k\not=1$, this is a contradiction with the fact
that the same ray has its turning points at A as well as B. To see this,
suppose that $k>1$ (in the opposite case we can relabel the points A and B), so
$r_{\rm B}>r_{\rm A}$. Then if the derivative $\dd N/\dd r$ is negative at
$r=r_{\rm A}$, then for the ray with angular momentum $L=N(r_{\rm A})$ the
region $r>r_{\rm A}$ is inaccessible because in this region $L>N(r)$, which is
impossible, and so the ray cannot make it to B. On the other hand, if the
derivative $\dd N/\dd r$ is positive at $r=r_{\rm A}$, then it is also positive
at $r=r_{\rm B}$. For a similar reason then the ray with angular momentum
$L=N(r_{\rm A})=N(r_{\rm B})$ cannot make it to A, which is a
contradiction. Hence the only possibility is $r_{\rm A}=r_{\rm B}$ which then
implies $k=1$.

We thus see that there are just two cases of imaging by spherically symmetric
absolute instruments: either the image is given by spherical inversion (since
$r_{\rm A}r_{\rm B}=k$) of the object, possibly combined with some rotation, or
it is congruent with it. We have arrived at a slightly stronger statement than
derived in~\cite{BornWolf} for absolute instruments in general.  We will see in
Sec.~\ref{method} that Eq.~(\ref{second}) corresponds to the generalised
Maxwell's fish eye; all other spherically symmetric absolute instruments
correspond to Eq.~(\ref{first}) with $k=1$, so they give images congruent with
the object and their magnification is unity. This means that the imaging by
such a device is perfect in the sense of geometrical optics~\cite{BornWolf}.

\section{Inverse scattering problem for spatially confined rays}
\label{inversion}

In the inverse scattering problem in mechanics, the task is to determine an
unknown potential from the scattering angle which is a known (e.g. measured)
function
of the impact parameter. The problem was solved for central potentials in 1953
by O. B. Firsov~\cite{Firsov1953} and the analogous problem in optics was
solved by K. Luneburg~\cite{Luneburg1964}. In the situations considered there,
the particle or light ray incides from infinity, undergoes scattering and
leaves for infinity again, so the motion is not spatially confined. However,
the situation where the motion is restricted to a finite region of space may be
equally important for design of absolute instruments and perfect lenses.  The
inverse problem can be formulated and solved also in this case; the solution
for mechanical motion was given without derivation by V. N. Ostrovsky
in~\cite{Ost97}.  In the following we derive the inversion formula for confined
motion in the optical case. 

To derive the inversion formula, we will assume that the function
$\rhho:r\mapsto\rhho(r)$ in Eq.~(\ref{rhhodef}) is increasing for
$r\!\leq\!r_{0}$ and decreasing for $r\!\geq\!r_{0}$ with some radius
$r_0>0$. (There are obviously more general profiles $\rhho(r)$ and the inverse
problem can be solved for them as well, but we will not
consider them here.) Then the function $\rhho(r)$ has a global maximum
$\ellzero\equiv\max\rhho\!=\!\rhho(r_0)$ at the point $r\!=\!r_{0}$. The
inverse function $r:\rhho\mapsto r(\rhho)$ is multi-valued, i.e., has two
branches $r_{\pm}$. One branch $r_{-}:\rhho\mapsto r_{-}(\rhho)$ maps into the
inner region $[0,r_{0}]$, and the other branch $r_{+}:\rhho\mapsto
r_{+}(\rhho)$ maps into the outer region $[r_0,\infty)$.  In this case we can
    invert the inequality $L\!\leq\!\rhho(r)$ into a double inequality
\begin{equation}
r_{-}(L)\leq r\leq r_{+}(L)\,,
\label{rhorplus}\end{equation}
which explicitly specifies the allowed confined region. In other words, for
angular momentum $L<\ellzero$, there are two turning points $r_{\pm}(L)$. 
Moreover, a light ray with angular momentum $L=\ellzero$ will propagate 
along a circular trajectory with radius $r_{0}$, while angular momentum 
$L>\ellzero$ is forbidden.

For the subsequent calculations, it turns out to be more convenient to work in
terms of a new coordinate $x=\ln r$ rather than the radius $r$. We will also
introduce corresponding notation $x_{\pm}=\ln r_{\pm}$, $x_{0}=\ln r_{0}$,
etc., in the obvious fashion.  At a general point of a ray trajectory, the
derivative of the polar angle $\fii$ is
\begin{equation}
\frac{\dd\fii}{\dd x}=r\frac{\dd\fii}{\dd r}=\tan\alpha
=\frac{L}{\sqrt{\rhho^{2}(r)-L^{2}}}\,,
\label{dfidr}\end{equation}
where we have used that $L\!=\!\rhho(r)\sin\alpha$. With the help of 
Eq.~(\ref{dfidr}), the increment of the polar angle corresponding to motion 
between $r_{-}(L)$ and $r_{+}(L)$ can be written as
\begin{equation}
\Delta\fii(L)=L\int_{r_{-}(L)}^{r_{+}(L)}\frac{\dd r}{r\sqrt{\rhho^{2}(r)-L^{2}}}
 =L\int_{x_{-}(L)}^{x_{+}(L)}\frac{\dd x}{\sqrt{\rhho^{2}(x)-L^{2}}}\,.
\label{Deltay}\end{equation}
We shall call $\Delta\fii$ the turning angle; it expresses the change of the
ray direction between two consequent turning points. 

\begin{figure}[htb]
\begin{center}
\includegraphics[width=90mm, angle=0]{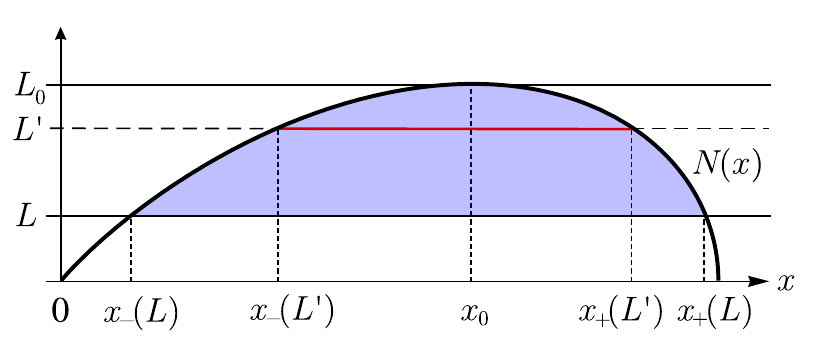}
\end{center}
\caption{The change of integration limits in Eq.~(\ref{xi}) illustrating that
  $\int_{L}^{\ellzero}\int_{x_{-}(\kay^{\prime})}^{x_{+}(\kay^{\prime})}u(x,L')\,\dd
  x \,\dd L' =\int_{x_{-}(L)}^{x_{+}(L)}\int_{L}^{\rhho(x)}u(x,L')\,\dd L'
  \,\dd x$.}
\label{limits}
\end{figure}

The task of the inverse scattering problem is to find the refractive index 
$n(r)$ [or equivalently, the function $\rhho(x)$] from the known turning 
angle $\Delta\fii(L)$. It can be solved in a similar way to finding the 
1D potential from the known period of oscillations as a function of energy 
\cite{Landau}. We divide $\Delta\fii(\kay^{\prime})$ by
$\sqrt{\kay^{\prime 2}-L^{2}}$, where $\kay^{\prime}$ is an integration
parameter, and integrate with respect to $\kay^{\prime}$ from $L$ to $\ellzero$:
\begin{align}\nonumber
\int_{L}^{\ellzero}
\frac{\Delta\fii(\kay^{\prime})\,\dd \kay^{\prime}}{\sqrt{\kay^{\prime 2}-L^{2}}}
\stackrel{(\ref{Deltay})}{=}&
\int_{L}^{\ellzero}\int_{x_{-}(\kay^{\prime})}^{x_{+}(\kay^{\prime})}
\frac{\dd x}{\sqrt{\rhho^{2}(x)-\kay^{\prime 2}}}
\frac{\kay^{\prime}\,\dd \kay^{\prime}}{\sqrt{\kay^{\prime 2}-L^{2}}} \\\nonumber
= & \int_{x_{-}(L)}^{x_{+}(L)}\int_{L}^{\rhho(x)}
\frac{\kay^{\prime}\,\dd \kay^{\prime}}{\sqrt{\kay^{\prime 2}-L^{2}}}
\frac{\dd x}{\sqrt{\rhho^{2}(x)-\kay^{\prime 2}}}\cr
=&\int_{x_{-}(L)}^{x_{+}(L)}\left[
\arcsin\sqrt{\frac{\kay^{\prime 2}-L^{2}}{\rhho^{2}(x)-L^{2}}}\,
\right]_{\kay^{\prime}=L}^{\kay^{\prime}=\rhho(x)}\dd x 
 \\ =&\int_{x_{-}(L)}^{x_{+}(L)}\left(\frac{\pi}{2}-0\right) \dd x 
= \frac{\pi}{2}\left(x_{+}(L)-x_{-}(L) \right)\,,
\label{xi}
\end{align}
where we have inverted the order of integration and changed the integration 
limits appropriately, see Fig.~\ref{limits}.
Equation~(\ref{xi}) yields the inversion formula
\begin{align}
\ln\frac{r_{+}(L)}{r_{-}(L)}= 
\frac{2}{\pi} \int_{L}^{\ellzero}
\frac{\Delta\fii(\kay^{\prime})\,\dd\kay^{\prime}}{\sqrt{\kay^{\prime 2}-L^{2}}}
\,,
\label{r+r-}\end{align}
which solves the inverse scattering problem. The mechanical equivalent of
Eq.~(\ref{r+r-}) was presented in \cite{Ost97} without derivation.

\section{Method for designing absolute instruments}
\label{method}

In the following we will describe a general method for designing absolute
instruments.  This method was sketched in~\cite{Ost97} but was not given
explicitly.

Consider the situation when the turning angle $\Delta\fii$ is independent of
$L$ and equal to $\pi/m$, $m\in\mathbb R$. This is a practically important
case; if, for instance, $m=p/q$ with $p,q$ coprimes and $p$ even, then the
trajectory will be symmetric with respect to rotation by $\pi$ around the
origin in the plane of propagation. A strong stigmatic image of a point A at
the position $\vec r$ will therefore be formed at $-\vec r$.  If $m=p/q$ with
$p$ odd, a strong stigmatic image of a point A will be formed at A itself; the
ray arrives there after encircling the origin $q$ times. This will be shown in
Fig.~\ref{lomenacara-fig}.

For $\Delta\fii(L)=\pi/m$ we get from Eq.~(\ref{r+r-})
$\ln[r_{+}(L)/r_{-}(L)]=(2/m)\,\arcosh(L_{0}/L)$, which can be expressed as
\begin{equation}
   \frac{L_{0}}L = \frac12\left[\left(\frac{r_{+}(L)}{r_-(L)}\right)^{m/2} +
   \left(\frac{r_{-}(L)}{r_{+}(L)}\right)^{m/2}\right].
\label{LmL}\end{equation}
Now comes the key step of our derivation. We define a function $f(r)$ such that
\begin{equation}
 f(r)=\begin{cases} r_+(\rhho(r)) \quad & \mbox{for}\quad r\le r_0 \\
    r_-(\rhho(r)) \quad & \mbox{for}\quad r\ge r_0\,.
\end{cases}
\label{deff}\end{equation}
The function $f(r)$ is hence  defined such that for a given lower turning point
$r_{-}$ it produces the upper turning point $r_+$ corresponding to the same
angular momentum and vice versa:
\begin{equation}
  r_{\pm}=f(r_{\mp})\,.
\label{f}\end{equation}
We can say that the point $r_+$ is dual to the point $r_-$ and vice versa.  It
also follows from this definition that applying the function $f$ twice yields
the original value, $f(f(r))=r$, and therefore the graph of $f$ is symmetric
with respect to the axis of the first quadrant. The graph intersects this axis
at the point $r_{+}=r_{-}=r_{0}$, which corresponds to the circular ray
trajectory.

With the function $f(r)$ defined this way, we can express $L$ from
Eq.~(\ref{LmL}) and either keep $r_-$ and replace $r_+$ by $f(r_-)$, or keep
$r_+$ and replace $r_-$ by $f(r_+)$. Taking then advantage of the fact that
$L=\rhho$ at $r_\pm$ and omitting the index of $r_\pm$, we get
\begin{equation}
  \rhho(r)=2L_{0}\left[\left(\frac{r}{f(r)}\right)^{m/2} 
     + \left(\frac{f(r)}{r}\right)^{m/2}\right]^{-1} \,.
\label{}\end{equation}
Finally we express $n(r)=\rhho(r)/r$ as
\begin{equation}
  n(r)=\frac{2L_{0}}{r\left[\left(\frac{r}{f(r)}\right)^{m/2} 
     + \left(\frac{f(r)}{r}\right)^{m/2}\right]}\,.
\label{nr}\end{equation}

Equation~(\ref{nr}) provides a powerful tool for designing absolute
instruments. For any chosen function $f(r)$ satisfying the condition above and
a suitable value of $m$ it gives a refractive index profile with focusing
properties, i.e., an absolute instrument. We will demonstrate this on several
known examples first and then proceed to new devices.

\section{Examples of absolute instruments}
\label{examples}

In all the following examples we assign unit radius and unit refractive index
to the circular ray, so $r_0=L_0=1$. In the first example the object and its
image are related by spherical inversion, in the other ones they are congruent,
as discussed in Sec.~\ref{theorem}.

\begin{figure}[htb]
\begin{center}
\begin{tabular}{ccc}
\includegraphics[height=65mm, angle=0]{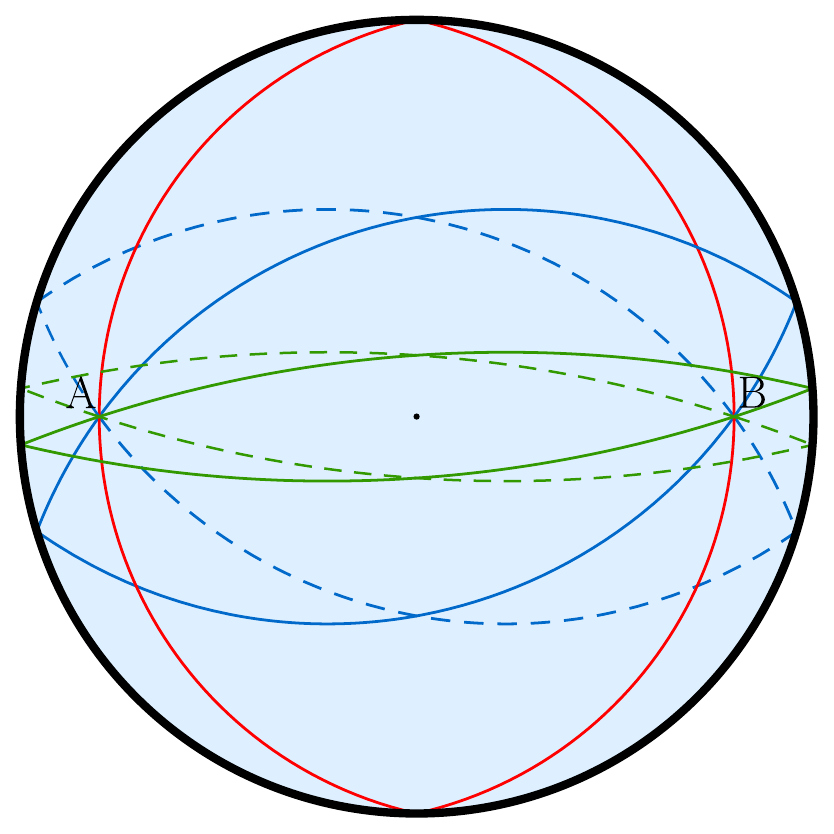}\hspace{2mm}&\hspace{5mm} &
\includegraphics[height=65mm, angle=0]{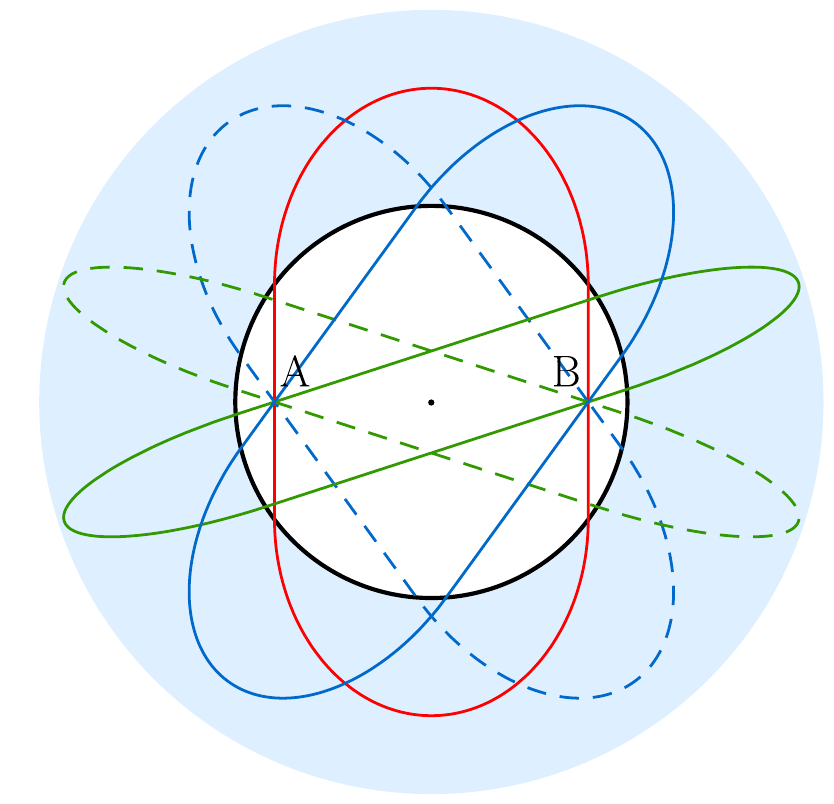}\hspace{2mm}\\
(a)&\hspace{5mm} & (b) \end{tabular}
\end{center}
\caption{Ray trajectories in (a) Maxwell's fish eye mirror an (b) Mi\~nano
  lens. Optically homogeneous region is shown in white colour. }
\label{mfe+minano}
\end{figure}

\begin{figure}[htb]
\begin{center}
\begin{tabular}{ccc}
\includegraphics[height=55mm, angle=0]{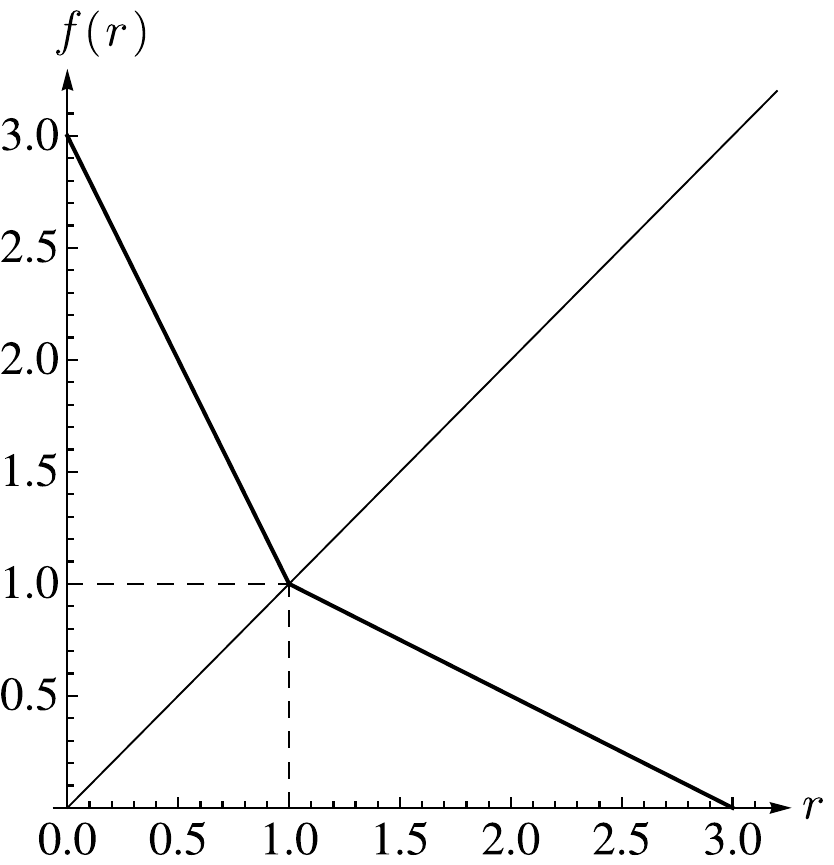}\hspace{2mm}&\hspace{5mm} &
\includegraphics[height=55mm, angle=0]{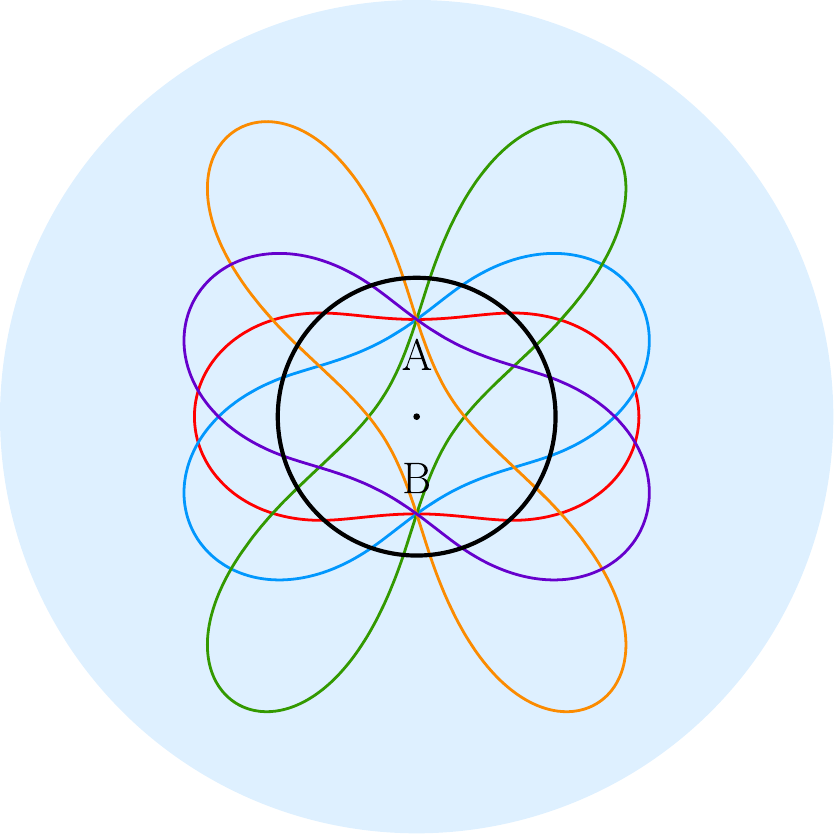}\hspace{2mm}\\
(a)&\hspace{5mm} & (b) \\[3mm]
\includegraphics[height=55mm, angle=0]{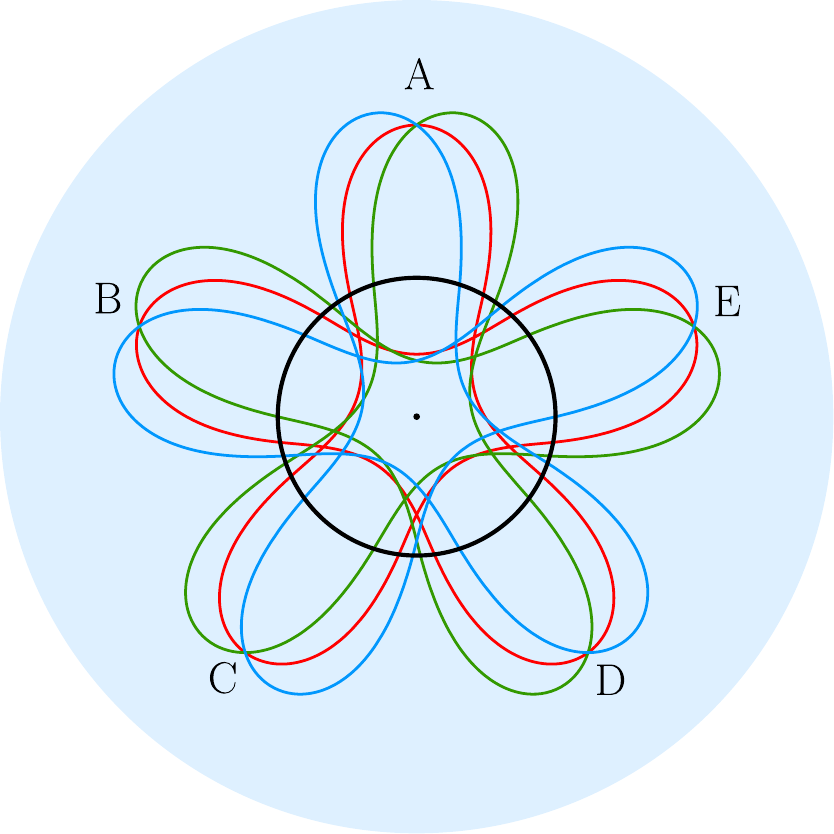}\hspace{2mm}&\hspace{5mm} &
\includegraphics[height=55mm, angle=0]{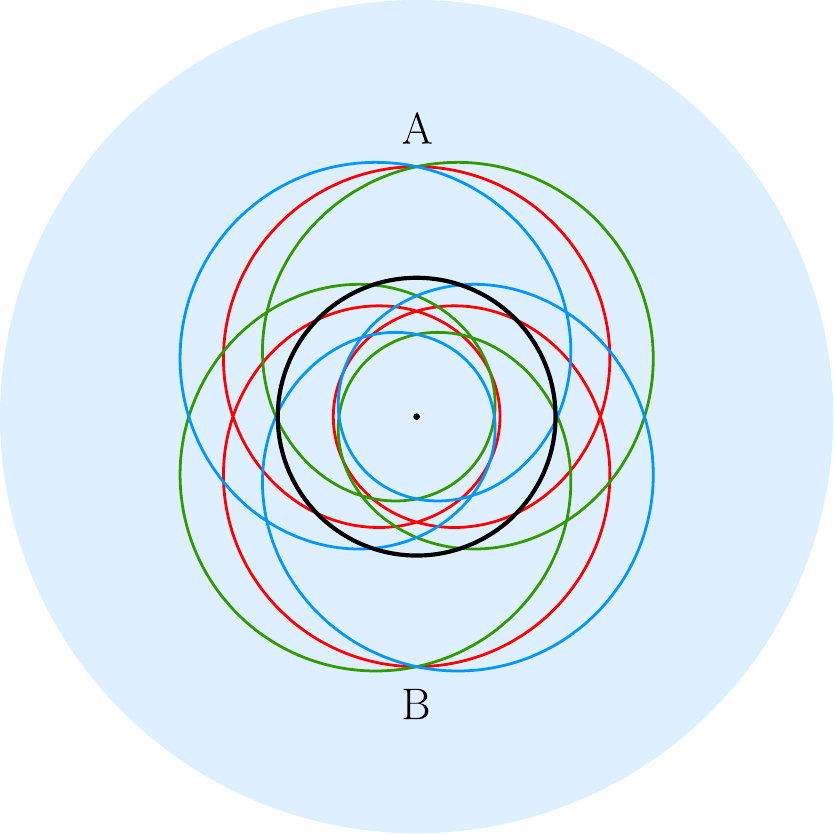}\hspace{2mm}\\
(c)&\hspace{5mm} & (d) \end{tabular}
\end{center}
\caption{Graph of a sample function $f(r)$ according to Eq.~(\ref{lomena-eq})
  and the corresponding ray trajectories for (b) $m=2$, (c) $m=5/2$ and (d)
  $m=2/3$. In (b) and (d), there is a strong image of A at B while in (c) the
  images B, C, D, E of A are just weak.}
\label{lomenacara-fig}
\end{figure}

\begin{itemize}
\item {\em Generalised Maxwell's fish eye}

Consider an absolute instrument in which the position of object and image is
given by Eq.~(\ref{second}). Apparently, if $r_{\rm A}$ is a turning point,
then $r_{\rm B}$ must also be a turning point, which leads to
$r_+r_-=k=r_0^2=1$ (the case of a general $k$ can be obtained easily by spatial
scaling) and hence $f(r)=1/r$.  Then Eq.~(\ref{nr}) yields
\begin{equation}
n(r)=\frac{2}{r^{1-m}+r^{m+1}}\,,
\label{gmfe}\end{equation}
which is the generalised Maxwell's fish eye profile discussed
in~\cite{Demkov1971}. Light rays in this lens are shown in
Fig.~\ref{fig-fisheye} (a) for $m=1/2$.

\begin{figure}[htb]
\begin{center} \begin{tabular}{ccc}
\includegraphics[height=65mm, angle=0]{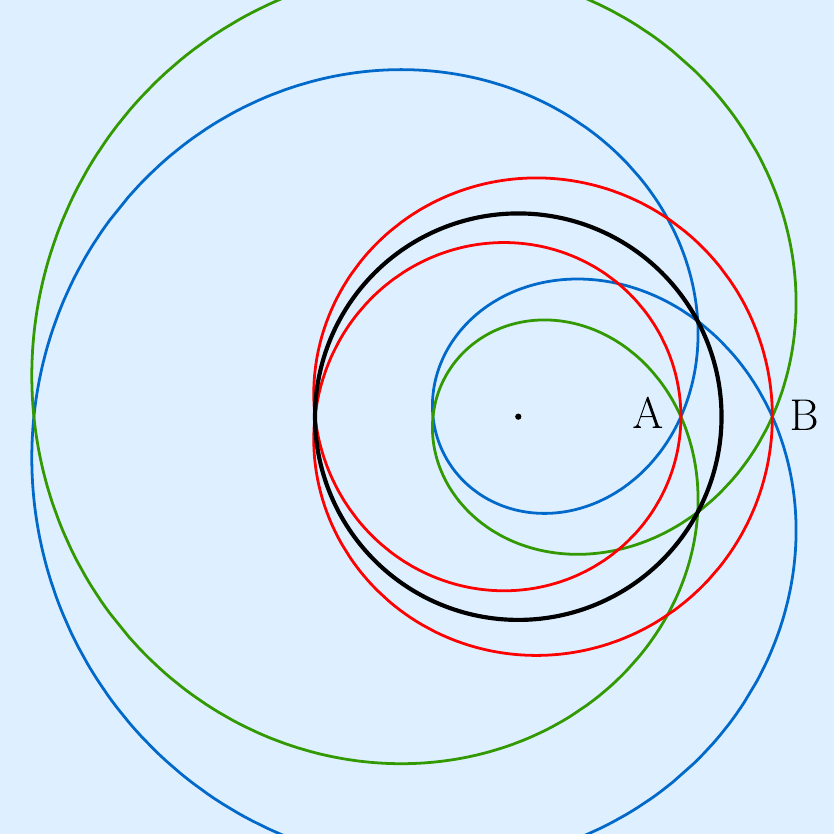} &\hspace{5mm} &
\includegraphics[height=65mm, angle=0]{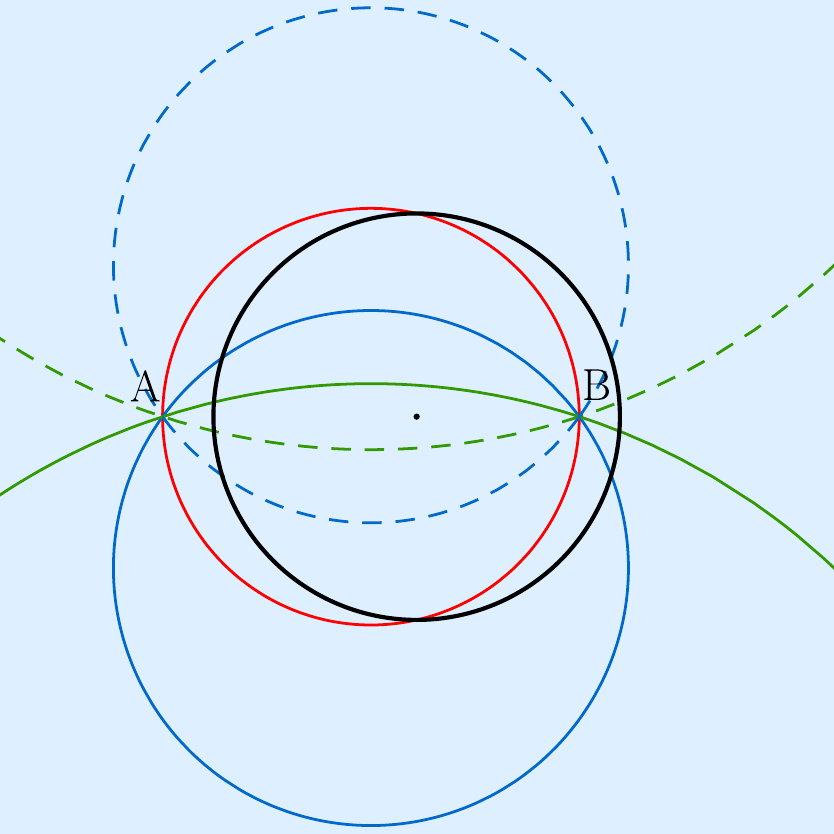}\\
(a)&\hspace{5mm} & (b) \end{tabular}
\end{center}
\caption{Ray trajectories in (a) generalised Maxwell's fish eye with $m=1/2$
  and (b) Maxwell's fish eye. The source and its image are denoted by A and B,
  respectively. Here as well as in all subsequent figures, the black circle
  corresponds to the circular trajectory at $r=r_0$ with the maximum possible
  angular momentum $L_0$.}
\label{fig-fisheye}
\end{figure}

For $m=1$ we get  
\begin{equation}
  n(r)=\frac{2}{1+r^2}\,,
\label{mfe}\end{equation}
which is the well-known Maxwell's fish eye refractive index
profile~\cite{MFE}. The trajectories are circles intersecting the unit circle
at two opposite points. Light rays are shown in Fig.~\ref{fig-fisheye} (b).

\item {\em Luneburg lens profile}

Take $m=2$ and $f(r)=\sqrt{2-r^2}$. Then Eq.~(\ref{nr}) yields
\begin{equation}
  n(r)=\sqrt{2-r^2}\,.
\label{V-Hooke}\end{equation}
For $r\le1$ this coincides with the well-known refractive index of Luneburg
lens~\cite{Luneburg1964} which has, however, $n=1$ for $r>1$.  The
index~(\ref{V-Hooke}) corresponds to Hooke potential in mechanics and ray
trajectories are ellipses centred at the origin \cite{Ulf-Thomas-book}, see
Fig.~\ref{fig-luneburg} (a).

\item {\em Eaton/Mi\~nano lens profile}

Take $m=1$ and  $f(r)=2-r$.  Then Eq.~(\ref{nr})
yields
\begin{equation}
   n(r)=\sqrt{(2/r)-1}\,.
\label{newton}\end{equation}
This is the well-known index profile of Eaton or Mi\~nano
lens~\cite{Eaton1952,Minano2006} which have, however, $n=1$ for $r>1$ or $r<1$,
respectively. The index (\ref{newton}) corresponds to elliptic motion in Newton
potential in mechanics and ray trajectories are confocal ellipses with focus at
the origin and with the main semiaxes of unit length \cite{Ulf-Thomas-book},
see Fig.~\ref{fig-luneburg} (b).

\begin{figure}[htb]
\begin{center}
\begin{tabular}{ccc}
\includegraphics[height=65mm, angle=0]{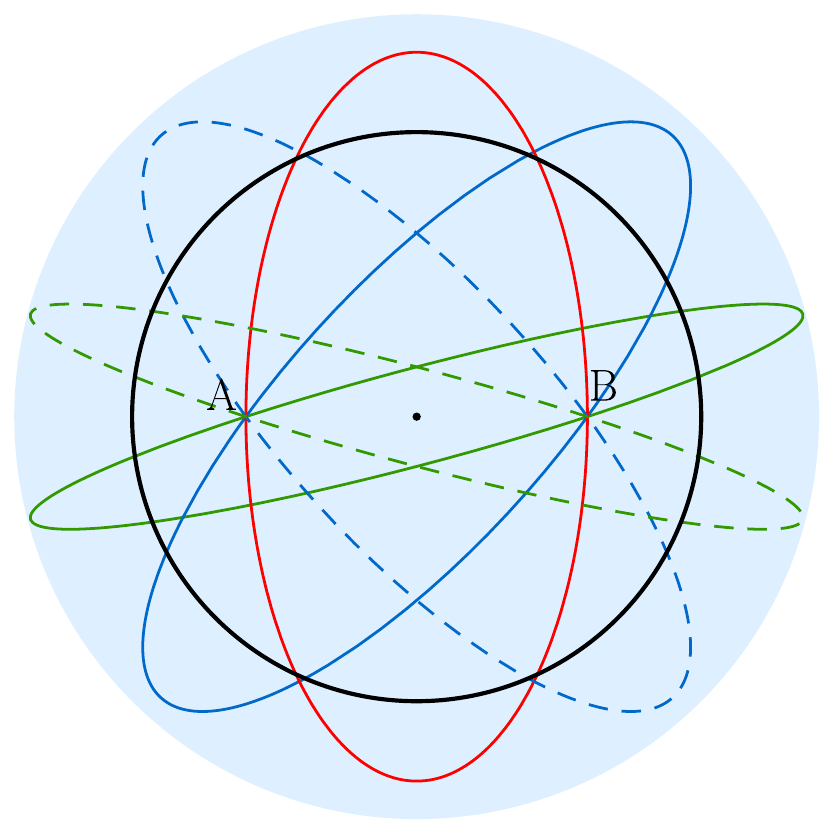}&\hspace{5mm} &
\includegraphics[height=65mm, angle=0]{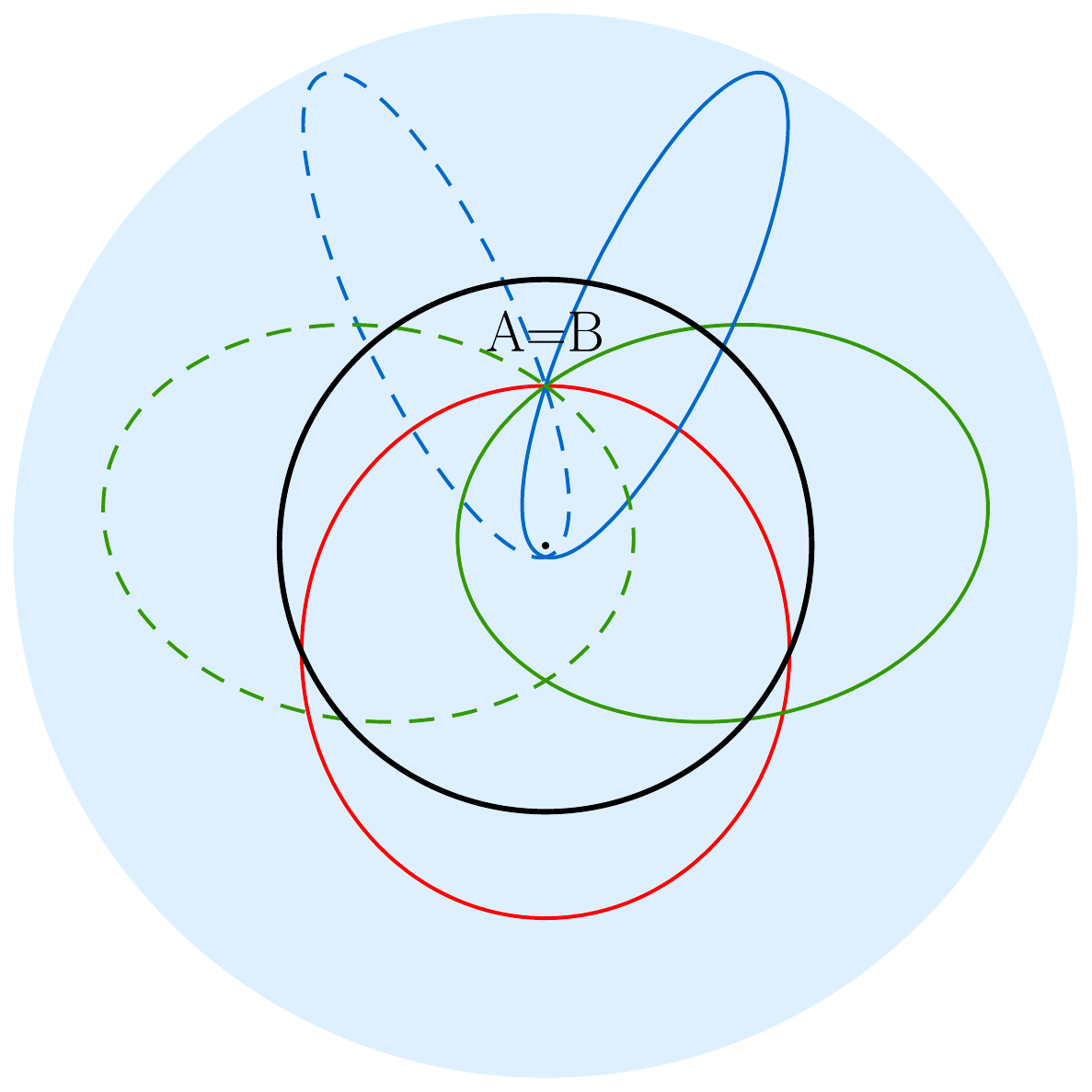}\\
(a)&\hspace{5mm} & (b) \end{tabular}
\end{center}
\caption{Ray trajectories in the profile given by (a) Eq.~(\ref{V-Hooke})
  corresponding to Luneburg profile and (b) Eq.~(\ref{newton}) corresponding to
  Eaton/Mi\~nano profile. The optical medium is shown in light blue colour; the
  refractive index goes to zero at the edge of the medium. In Eaton/Mi\~nano
  profile the image B coincides with the source A.}
\label{fig-luneburg}
\end{figure}

\item {\em Maxwell's fish eye mirror}

Take $m=2$ and $f(r)=1$ for $r<1=r_0$. From the symmetry of the function $f$ it
follows that it is undefined for $r\ge1$; the rays cannot get beyond the unit
circle, which means they must be reflected there. Substituting into
Eq.~(\ref{nr}), we get for $r\in[0,1)$ the index given by Eq.~(\ref{mfe}).
This is the so-called  Maxwell's fish eye mirror discussed in detail
in~\cite{Ulf2009-fisheye} with rays shown in Fig.~\ref{mfe+minano}.

\item {\em Lenses generated by a sample function $f(r)$}

To show the generality of our method, we choose some arbitrary function $f(r)$
with the restriction that it is symmetric with respect to the axis of the first
quadrant. Let us choose
\begin{equation}
f(r)= \begin{cases} 3-2r  \quad & \mbox{for}\quad r\le 1\\
                    (3-r)/2 \quad & \mbox{for}\quad r\geq 1
\end{cases}
\label{lomena-eq}
\end{equation}
see its graph in Fig.~\ref{lomenacara-fig} (a). Using different values of $m$,
we get different absolute instruments. The rays in some of them are shown in
Fig.~\ref{lomenacara-fig} (b) -- (d) for $m=2$, $m=5/2$ and $m=2/3$.

\end{itemize}

\subsection{Lenses with mirrors and index discontinuities}
\label{mirrors}

On the example of Maxwell's fish eye mirror we have seen that an interval of
$r$ on which the function $f(r)$ from Eq.~(\ref{f}) was constant corresponded
to a spherical mirror. As we will see, this is quite a general feature.

Consider the situation when the function $f(r)$ is constant and equal to $c\ge
r_0$ on some interval $(a,b)$ with $b\le r_0$. In other words, the upper
turning point is the same for different lower turning points and hence for
different angular momenta; the corresponding rays cannot get beyond $r=c$,
which means there occurs reflection at this point.  Now there are two cases to
be distinguished. In the first case $f(r)=c$ holds also for $0\le r<a$, which
means that a reflection occurs for all rays that reach the point $r=c$; this
corresponds to a perfect mirror placed at $c$. In the second case, the value
$f(r)$ gets larger than $c$ for some $r<a$.  This means that rays with small
angular momenta can penetrate beyond $c$; the reflection at $c$ mentioned above
must then be total internal reflection caused by a refractive index
discontinuity at $r=c$. That such a discontinuity indeed exists follows from
the fact that if $f(r)=c$ on the interval $(a,b)$, then $f(r)$ must be
discontinuous at $r=c$ and so must be $n(r)$ according to Eq.~(\ref{nr}).

\begin{figure}[htb]
\begin{center}
\begin{tabular}{ccc}
\includegraphics[width=60mm, angle=0]{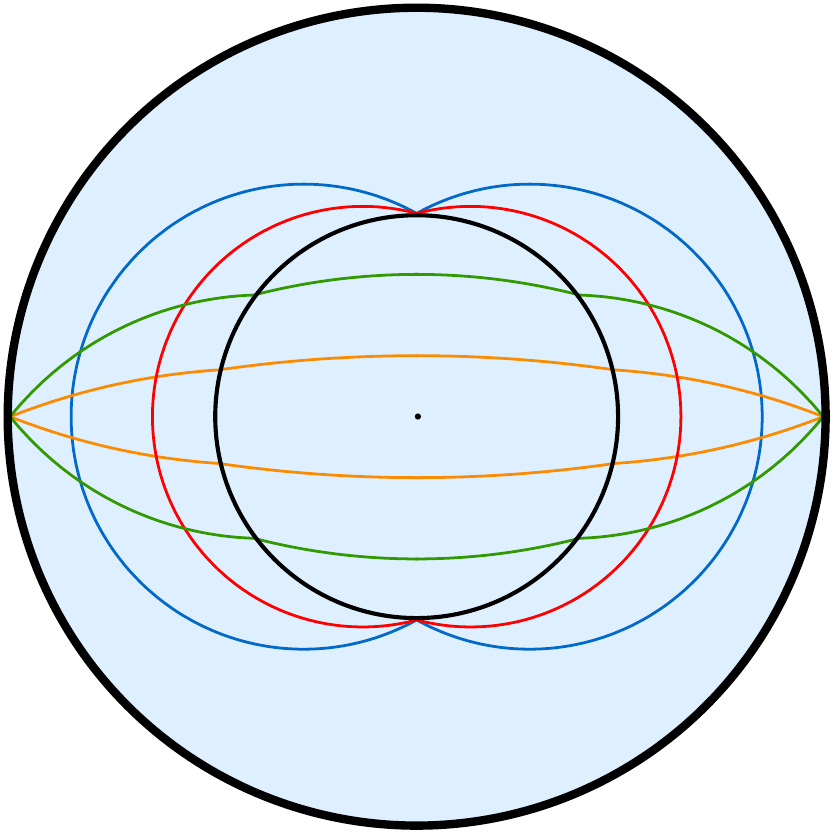}&\hspace{5mm} &
\includegraphics[width=60mm, angle=0]{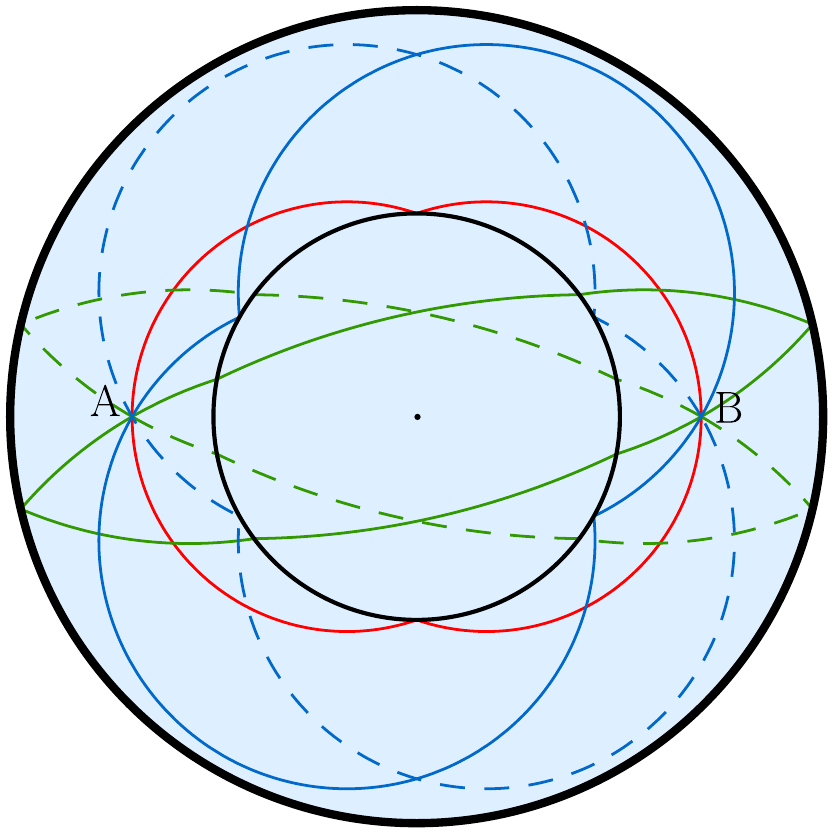}\\
(a)&\hspace{5mm} & (b) \\
\includegraphics[width=55mm, angle=0]{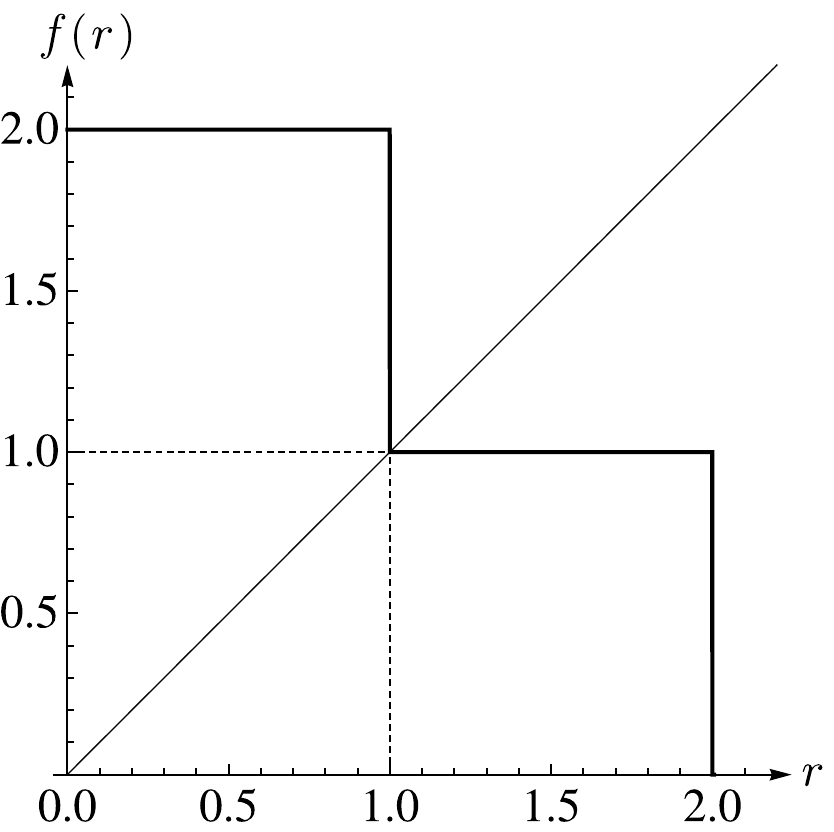}&\hspace{5mm} &
\includegraphics[width=55mm, angle=0]{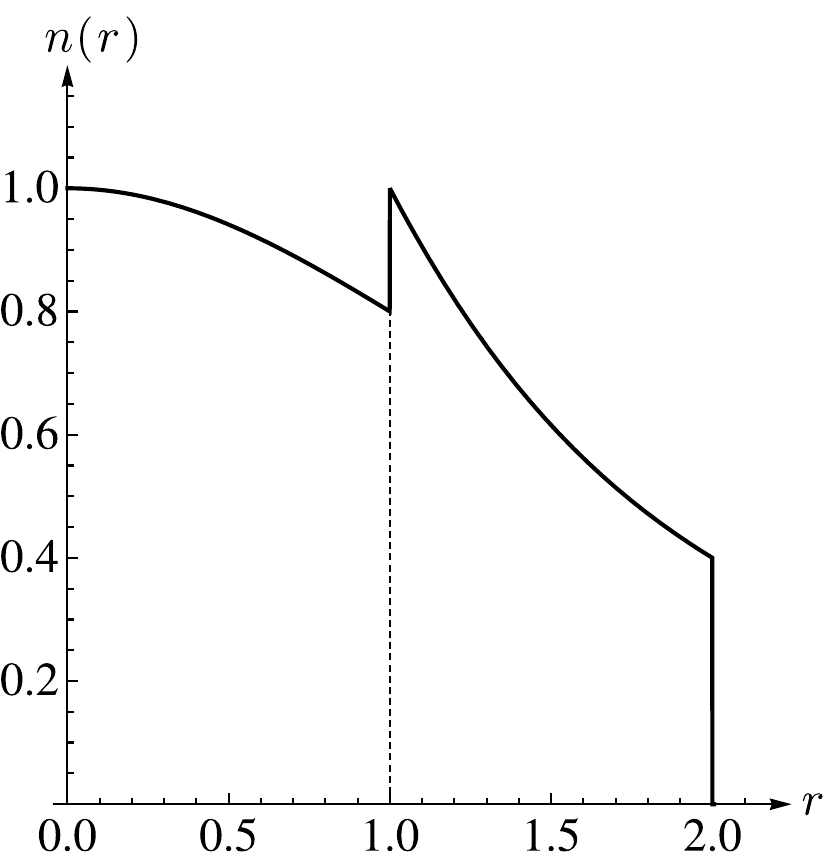}\\
(c)&\hspace{5mm} & (d) \\
\end{tabular}
\end{center}
\caption{(a) Ray trajectories and (b) imaging by an absolute
  instrument with a spherical mirror at $r=2$ and a jump of refractive index
  at radius $r=1$. There are two types of ray trajectories differing by whether
  they do or do not penetrate inside the unit circle. (c) The corresponding
  function $f(r)$ and (d) refractive index. }
\label{fig-reflections}
\end{figure}

If the interval of constant $f$ lies above $r_0$ then there will be a total
reflection on a sphere ``from the outside''.  This is illustrated in
Fig.~\ref{fig-reflections} where there is a mirror reflecting perfectly from
the inside and a jump of refractive index reflecting totally some rays from the
outside.

Spherical mirrors can be used with a great advantage to reduce the size of the
lens and also the range of the refractive index. An example is Maxwell's fish
eye mirror discussed above that has equally good focusing properties as the
original Maxwell's fish eye but its size is reduced from infinity to a unit
disc (or sphere) and the refractive index range is reduced significantly from
the interval $(0,2]$ to just $[1,2]$.

\section{Absolute instruments for homogeneous regions}
\label{homogeneous}

In many absolute instruments such as Maxwell's fish eye, the optical medium
fills the whole space and the object to be imaged must therefore be inserted
into such an inhomogeneous medium. It is desirable to find optical devices that
provide images of optically homogeneous spatial regions, i.e., regions with a
uniform refractive index. Even if the refractive index of such a region differs
from unity, this is still an advantage because one can fill this region with a
suitable liquid and place the object in it.

Interestingly, until recently the only known devices providing stigmatic images
of optically homogeneous 3D regions were plane mirrors or their
combinations~\cite{BornWolf}.  This has changed by an excellent work of
J. C. Mi\~nano~\cite{Minano2006} who noticed that some well-known optical
devices such as Eaton or Luneburg lens~\cite{Luneburg1964} are in fact absolute
instruments, providing stigmatic virtual images.  An even more important result
of~\cite{Minano2006}, however, is a a device now called Mi\~nano lens that
provides real images of homogeneous 3D regions.

\subsection{Mi\~nano lens}

In Mi\~nano lens the homogeneous region is the unit disc with unit refractive
index and the turning angle is $\Delta\fii=\pi/2$, which corresponds to $m=2$.
To derive the refractive index outside the unit disc by our method, we will
again use Eq.~(\ref{nr}), but first we have to determine the function $f$. The
lower turning point for a given $L$ is simply $r_-=L$ because $n=1$ in this
region, the upper turning point is $r_+=f(r_-)$. Substituting this into
Eq.~(\ref{LmL}), we get a quadratic equation $f^2(r_-)-2f(r_-)+r_-^2=0$ for
$f(r_-)$ with the solution
\begin{equation}
r_+=f(r_-)=1+\sqrt{1-r_-^2}
\label{minanof}\end{equation}
(we have taken the larger root so that $r_+\ge r_-$). Inverting this expression
to get the function $f$ also for $r_+$, we find $r_-=f(r_+)=\sqrt{2r_+-r_+^2}$.
Now we can combine Eqs.~(\ref{nr}) and~(\ref{minanof}) with $L_0=1$ to find the
refractive index outside the unit disc.  This gives precisely the
expression~(\ref{newton}), so the refractive index of Mi\~nano lens is
 \begin{equation}
n(r)=\begin{cases} 1 \quad & \mbox{for}\quad r\le 1 \\
   \sqrt{(2/r)-1}\quad & \mbox{for}\quad r>1
\end{cases}
\label{minano}\end{equation}
and the ray trajectories are shown in Fig.~\ref{mfe+minano} (b).

\subsection{Modified Mi\~nano lens}

\begin{figure}
 \begin{center} \begin{tabular}{ccc}
 \includegraphics[width=60mm]{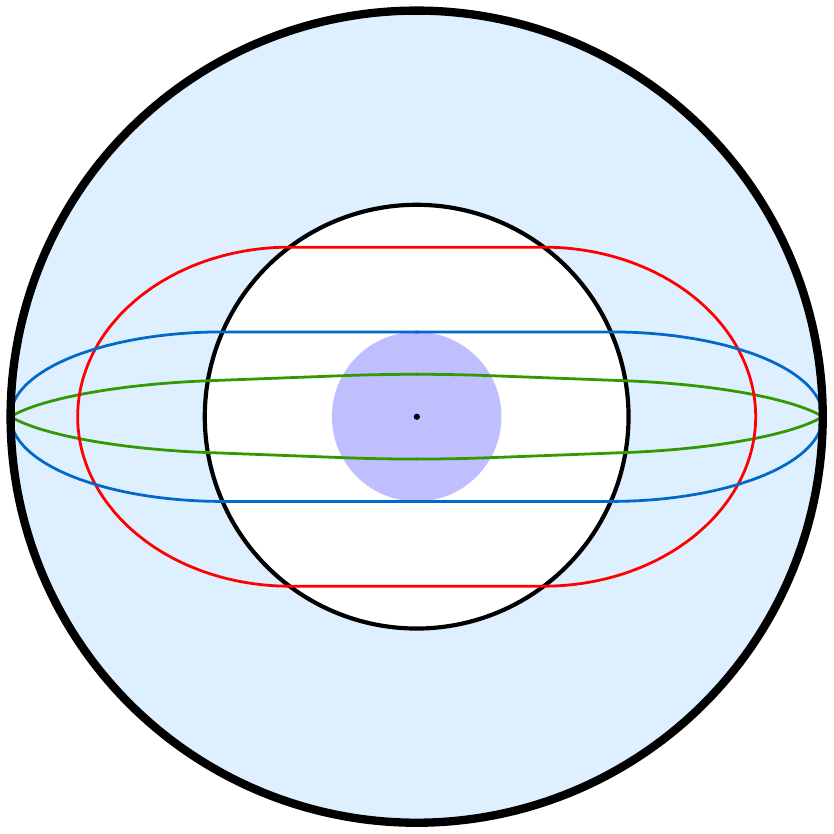} & \hspace{5mm}&
 \includegraphics[width=60mm]{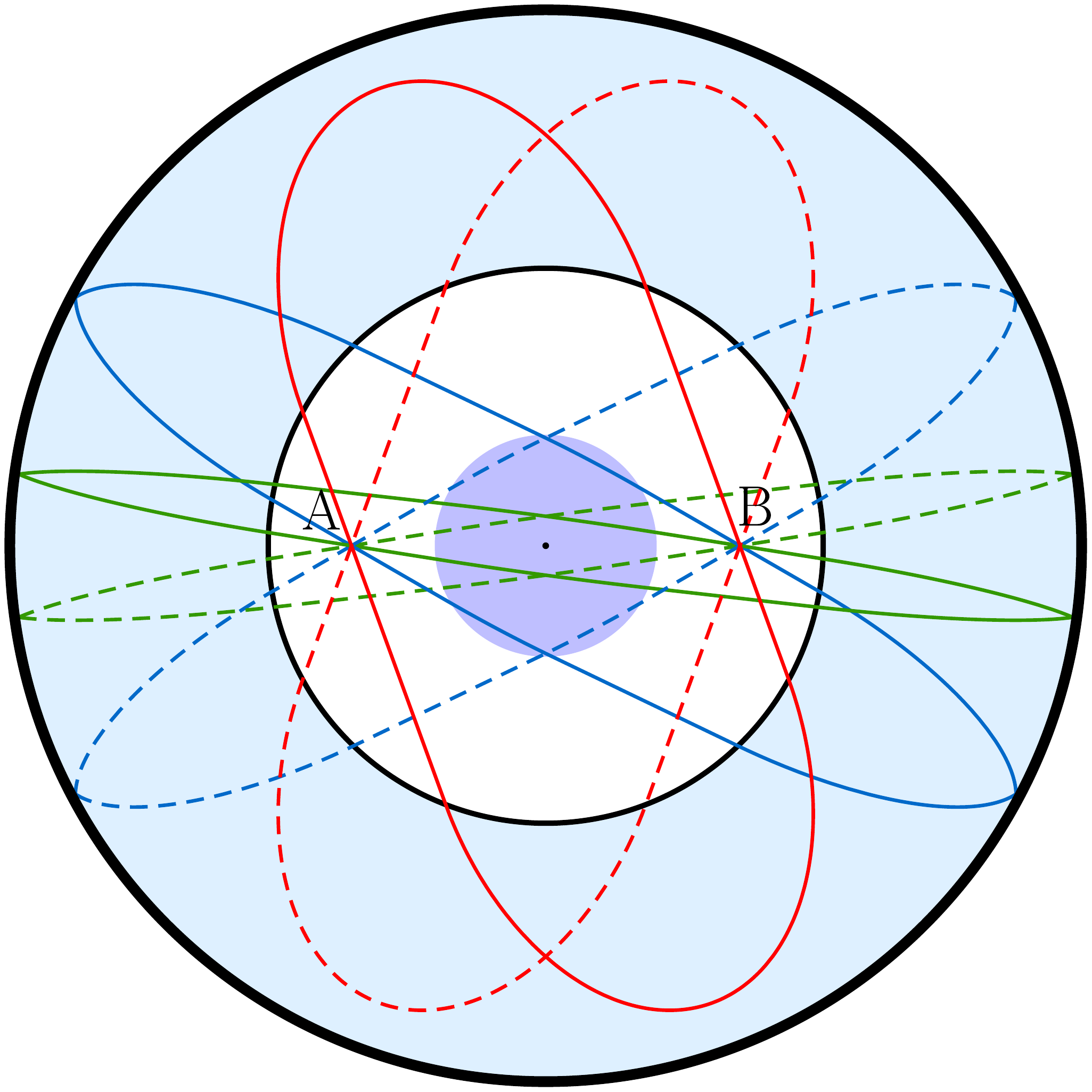}  \\
   a)& \hspace{5mm} & b)\\
 \includegraphics[width=55mm]{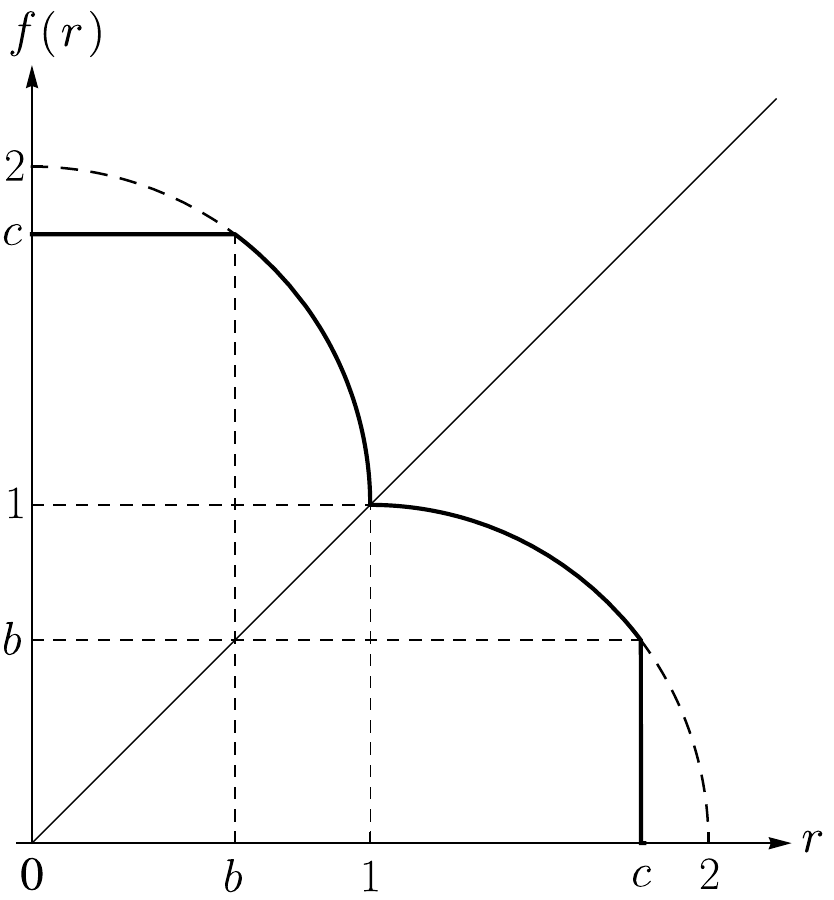} &  \hspace{5mm}&
 \includegraphics[width=55mm]{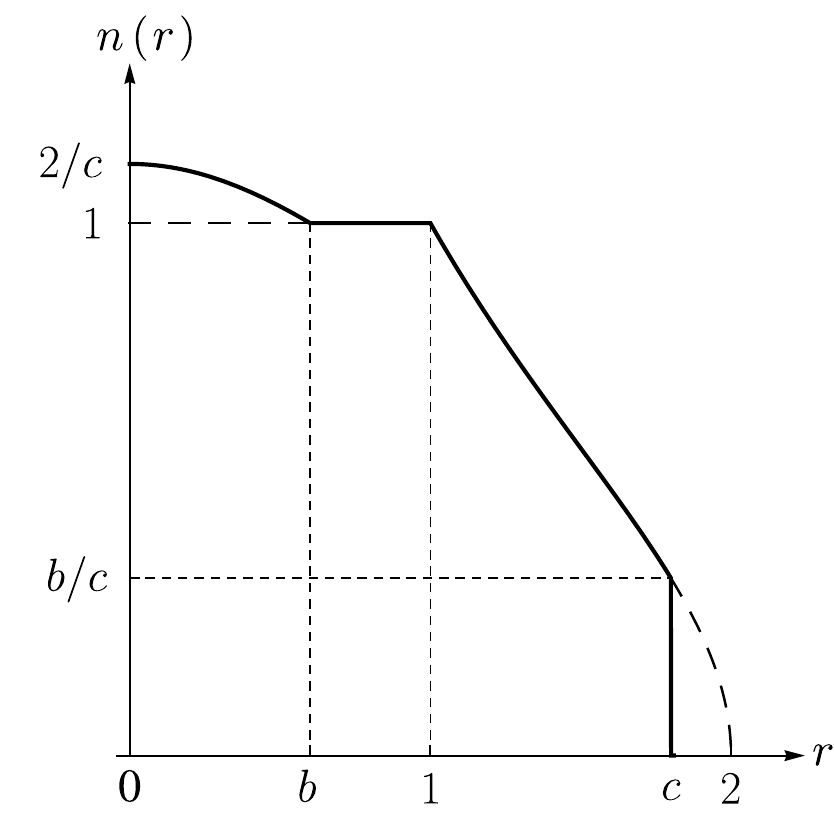}  \\
   c)& \hspace{5mm} & d)\\
      \end{tabular}\end{center}
 \caption{(a) Ray trajectories and (b) imaging by the modified Mi\~nano
   lens. In contrast to Mi\~nano lens, there is a spherical mirror at radius
   $c$ and an inner inhomogeneous region of radius $b$.  (c) The corresponding
   function $f(r)$ and (d) refractive index $n(r)$. The dashed parts of the
   graphs on the intervals $[0,b]$ and $[c,2]$ correspond to the original
   Mi\~nano lens. }
\label{minano-mod}
\end{figure}

Despite its elegance and nice properties, Mi\~nano lens has a disadvantage:
its refractive index ranges from unity all the way to zero at $r=2$, which is
very difficult to realise practically. Since any refractive index profile can
be multiplied by a real number without affecting the lens performance, we can
define a measure of the index range as the ratio of its largest and smallest
value
\begin{equation}
  \eta\equiv n_{\rm max}/n_{\rm min}\,.
\label{eta}\end{equation}
For Mi\~nano lens $\eta=\infty$.  It would be very desirable to modify the lens
somehow to make $\eta$ finite.  Fortunately, this is possible, as we will show
now. Imagine we use the function $f$ for $r<1$ according to
Eq.~(\ref{minanof}), but on the interval $[0,b]$ with $0<b<1$ we replace it by
a constant value $c=1+\sqrt{1-b^2}$. As we have seen, this corresponds to
placing a mirror at $r=c$. The function $f$ this way becomes
\begin{equation}\label{fm2}
 f(r)=\begin{cases} c \quad& \mbox{for} \quad r\leq b\\		
		1+\sqrt{1-r^2} \quad & \mbox{for}\quad b<r\leq 1\\
		\sqrt{2r-r^2}\quad & \mbox{for}\quad 1<r\leq c	 \end{cases}
\end{equation}
see Fig.~\ref{minano-mod} (c), and refractive index 
\begin{equation}\label{nm2}
 n(r)= \begin{cases}2c/\left(r^2+c^2\right) \quad & \mbox{for}\quad r\leq b\\ 1
   \quad & \mbox{for}\quad b<r\leq 1\\ \sqrt{2/r-1} \quad& \mbox{for}\quad
   1<r\leq c \end{cases}
\end{equation}
see Fig.~\ref{minano-mod} (d). The largest and smallest values of the index
occur at $r=0$ and $r=c$, respectively, and yield the ratio $\eta=2/b$.
Choosing $b$ not too small, one can get a very moderate index range of the
lens; the price to pay is that the size of the homogeneous region is reduced.
Rays in modified Mi\~nano lens and its imaging properties are shown in
Fig.~\ref{minano-mod} (a) -- (b).

\subsection{A lens for designing a magnifying absolute instrument}
\label{tomandmartin}

\begin{figure}
 \begin{center} 
 \includegraphics[width=65mm]{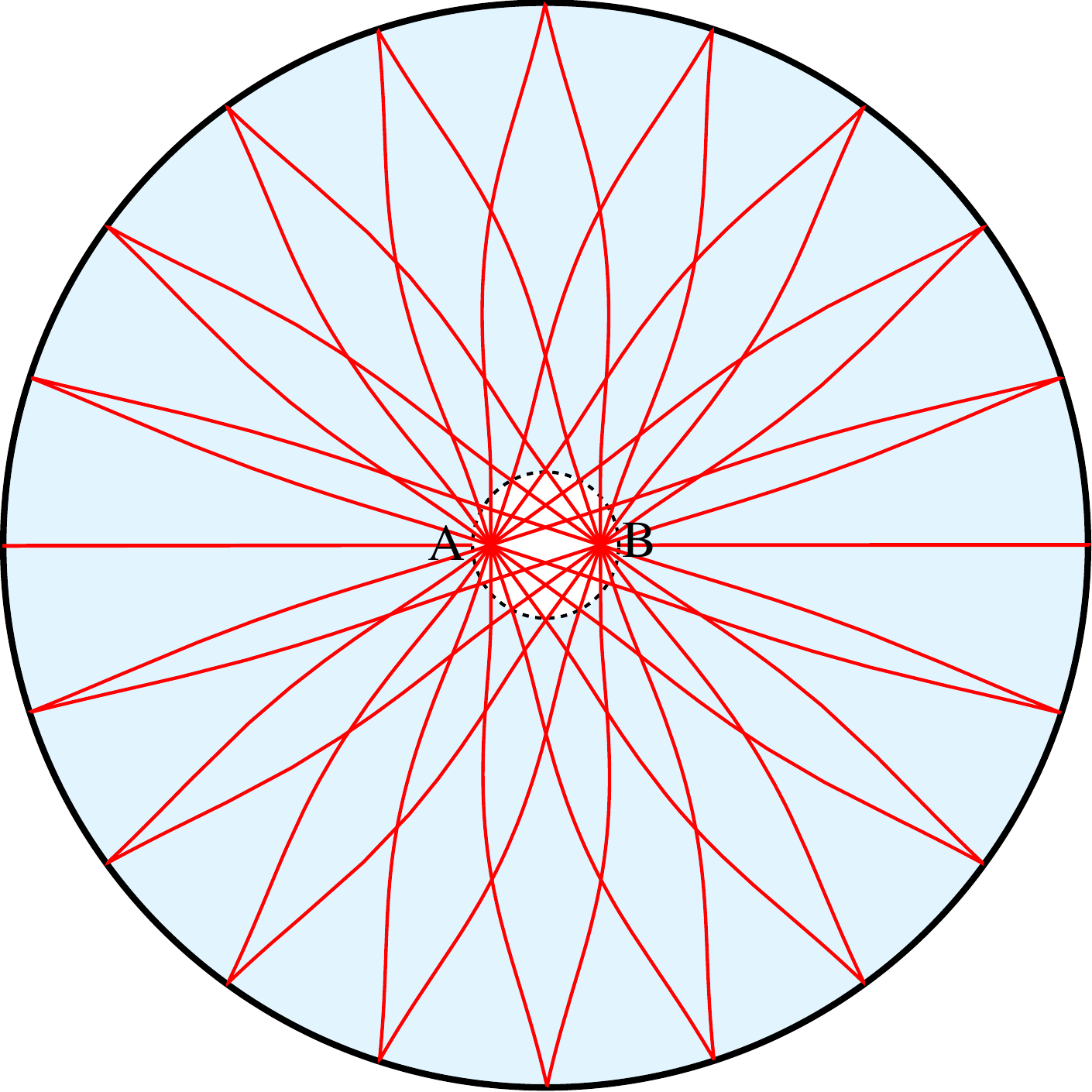}  
\end{center}
 \caption{Rays in the lens imaging a circular homogeneous region employing a
   spherical mirror. The homogeneous region is encircled by a dotted
   line. Unfortunately, the refractive index performing such an imaging
   perfectly does not exist as we show in Appendix.}
\label{nefunguje}
\end{figure}

Recently we proposed another type of lens for imaging homogeneous spatial
regions~\cite{Tyc2010} that has a finite ratio of the largest and smallest
refractive index. The homogeneous region is a unit sphere and there is a
spherical mirror at radius $R>1$. The refractive index between the two spheres
is chosen such that a ray emerging in the direction of a unit vector $\vec u$
from a point A located at $\vec r_{\rm A}$ in the homogeneous region incides on
the mirror at the point $R\vec u$. This ensures that the ray after the
reflection from the mirror passes through the point B at $\vec r_{\rm B}=-\vec
r_{\rm A}$ where an image of A is formed, see Fig.~\ref{nefunguje}.  Another
property of the device is that, apparently, all mutually parallel rays
propagating in the homogeneous region are focused to a single point at the
mirror.

The refractive index calculated for this lens was based on numerical solution
of a specific integral equation that does not have an analytic solution.
Unfortunately, it has turned out later that the equation does not have solution
at all, so the proposed lens is in fact not an absolute instrument. The rays
emerging from A hence do not pass exactly through B but there is some
unavoidable aberration. However, numerical simulations reveal that this
aberration can be very small if $R$ is not too small. For this reason the lens
could have a practical importance, and we are therefore mentioning it here
although it is not really an absolute instrument.  The proof of non-existence
of the solution of the integral equation is given in the Appendix.

\section{Lenses with multiple image points}
\label{multiple}

\begin{figure}[htb]
\begin{center} \begin{tabular}{ccc}
\includegraphics[height=65mm, angle=0]{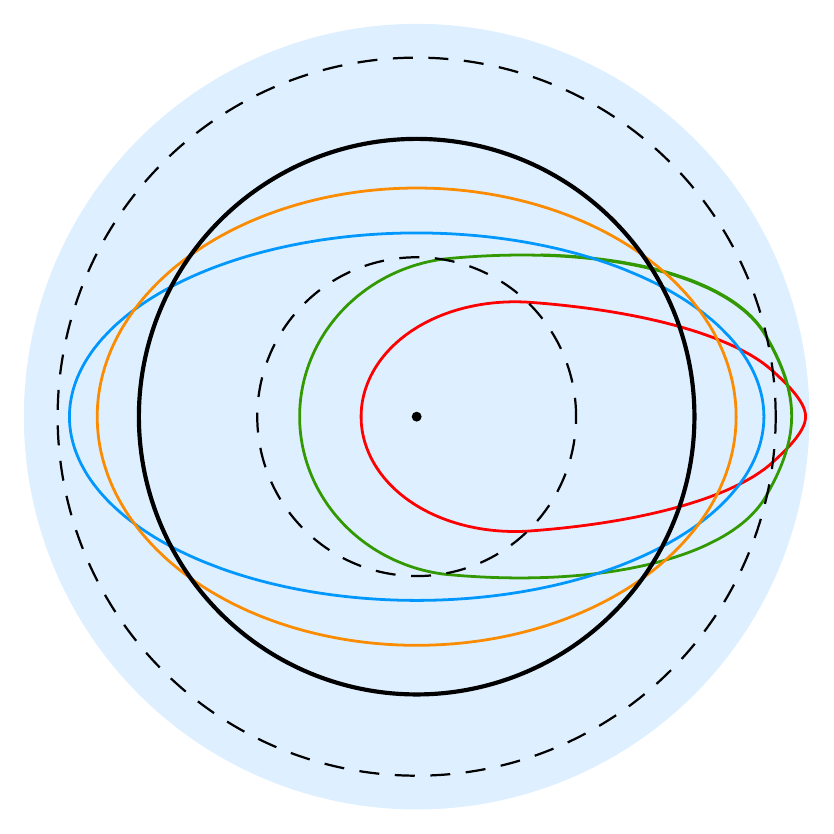}
& \hspace{5mm}&
\includegraphics[height=65mm, angle=0]{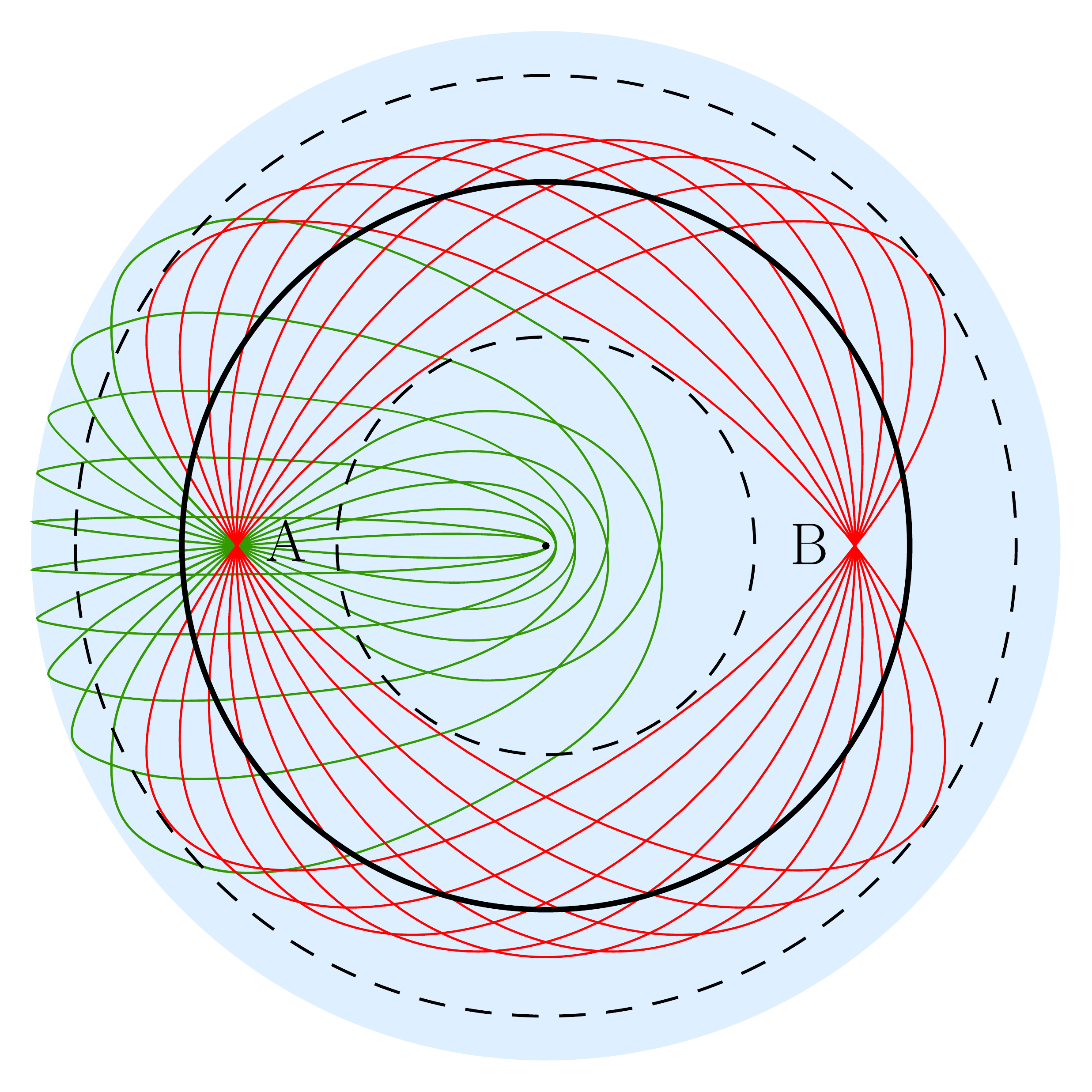}\\
  a)& \hspace{5mm} & b)\\
\includegraphics[height=65mm, angle=0]{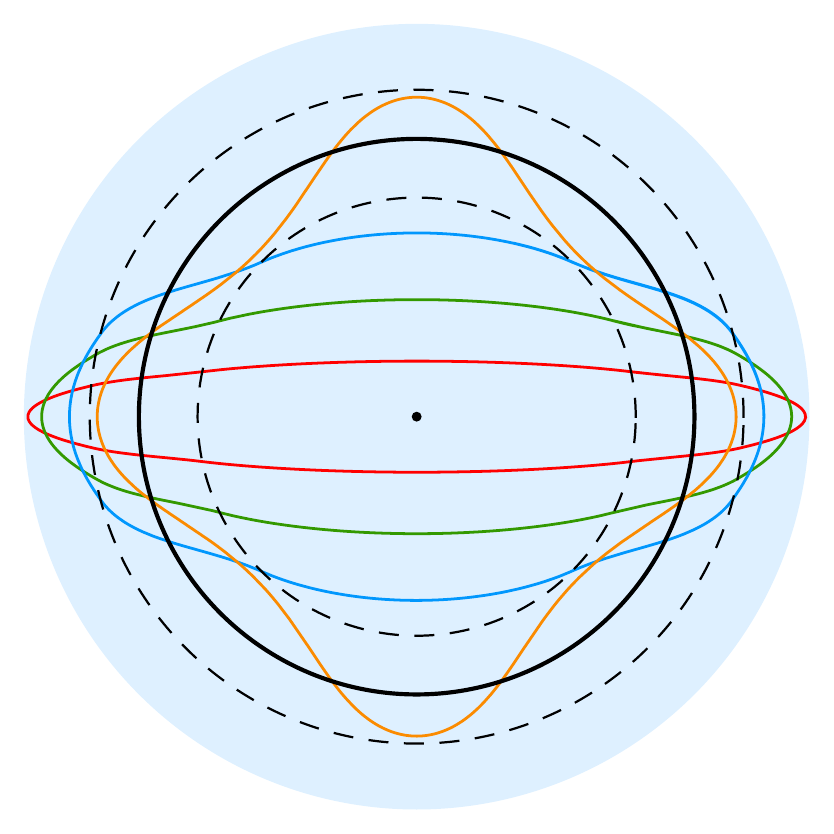}
& \hspace{5mm}&
\includegraphics[height=65mm, angle=0]{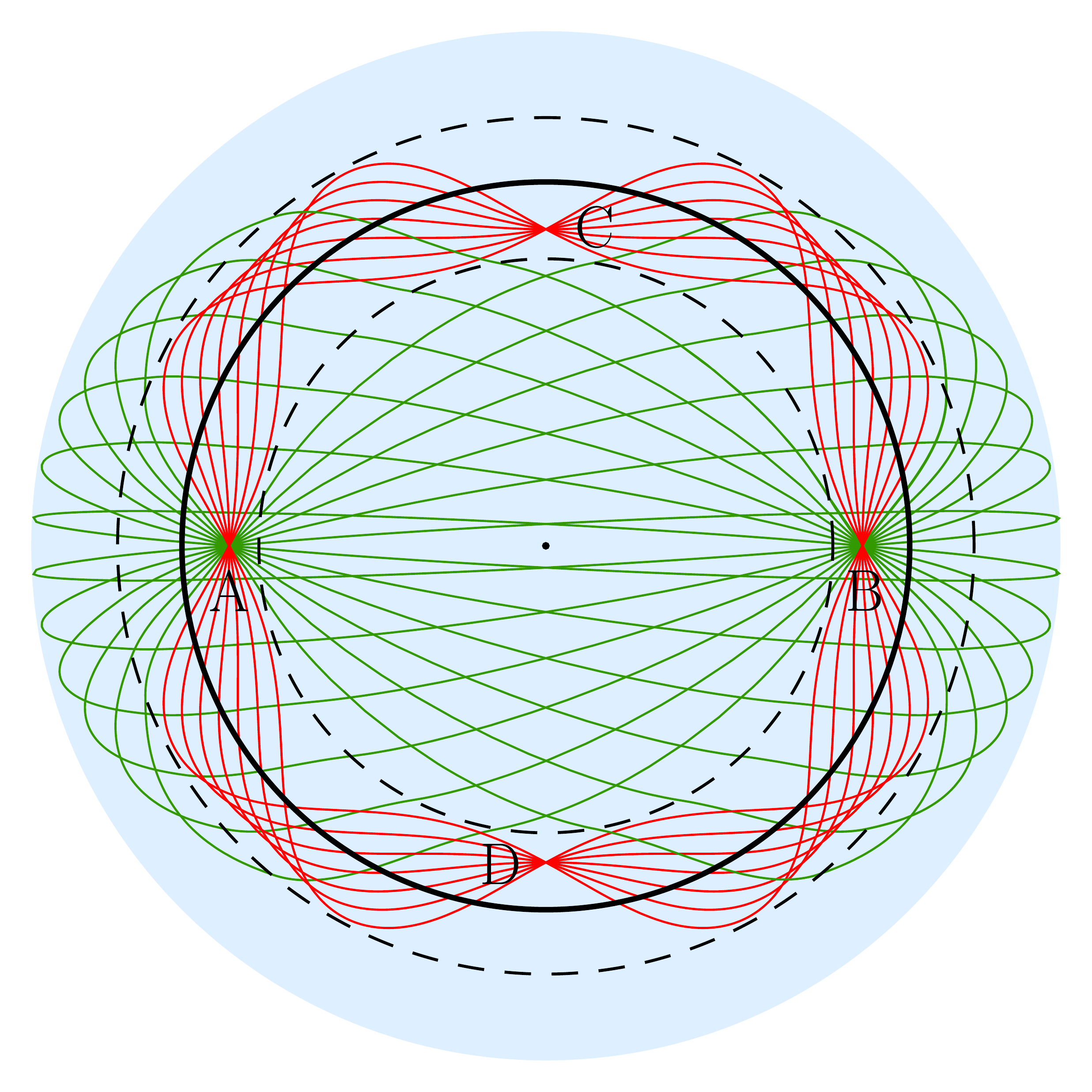}\hspace{2mm}\\
  c)& \hspace{5mm} & d)\\
\end{tabular}
\end{center}
\caption{Ray trajectories in ``bifocal'' lens with $L_0 = 1,L_1 = 3/4$,
  $f(r)=\sqrt{2a^2-r^2}$ and (a) -- (b) $m_1=1$, $m_2=2$ and (c) -- (d)
  $m_1=2$, $m_2=4$. The dashed circles mark the radii of turning points
  corresponding to angular momentum $L_1$. Rays confined to the ring between
  these circles have turning angle $\Delta\fii=\pi/m_2$ while rays that get outside the
  ring have $\Delta\fii=\pi/m_1$.}
\label{bifocal}
\end{figure}

So far, we have considered just situations when the turning angle
$\Delta\fii(L)$ in Eq.~(\ref{Deltay}) is constant for all angular momenta $L$
from zero to $L_0$. However, another class of interesting absolute instruments
can be obtained by taking $\Delta\fii(L)$ piecewise constant on different
intervals of $L$.  In this situation, however, the refractive index cannot be
found analytically with the help of Eq.~(\ref{r+r-}) but must be calculated
numerically.  The simplest situation corresponds to two values of turning
angle: $\Delta\fii(L)=\pi/m_1$ for $L\in[0,L_1]$ and $\Delta\fii(L)=\pi/m_2$
for $L\in[L_1,L_0]$.  Fig.~\ref{bifocal} (a) -- (b) shows an example of this
lens for $m_1=1$ and $m_2=2$.  Rays emerging from the point A with angular
momentum larger than $L_1$ meet first at the point B where they form an image
and then continue back to A.  On the other hand, rays with $L<L_1$ go back to A
directly without passing through B. This way, the image B is formed just by
some rays while the image A is formed by all rays.

Since there is not a smooth transition from the first class of rays (those with
$L>L_1$) to the second class ($L<L_1$), the optical path from A to B does not
have to be the same for rays of the two types.  In other words, the principle
of equal optical path~\cite{BornWolf} does not apply to this type of lens.  

Another example of such a ``bifocal lens'' is shown in Fig.~\ref{bifocal} (c)
-- (d) for $m_1=2$ and $m_2=4$. In the two-dimensional version of this problem,
some rays emerging from A form an image at C or D and then continue to the
image B. Other rays go to B directly without passing through C or D.  If the
lens is three-dimensional, a strong image will be just at B (and then at A
again), but weak images of A will be formed on the circle made by rotation of
the points C and D about the axis AB.  For coherent light waves emitted from
the source at A, the inequality of the optical paths from A to B for the two
ray types could lead to interesting diffraction patterns around the image point
B if the two corresponding waves interfere destructively.

\section{Conclusion}
\label{conclusion}

In this paper we have derived a number of results for imaging by spherically
symmetric absolute instruments. We have shown that images created by such
instruments are either congruent with the object or related to it by spherical
inversion. We have also proved that the mutual position of a point and its
strong image is quite restricted in 3D: the straight line connecting the two
points must intersect the centre of the lens.

Further, we have developed a general method for designing absolute instruments
via solving the inverse problem for finite motion.  We have shown how this
method can be used for designing already known as well as a number of new
absolute instruments.  The modified Mi\~nano lens we have proposed is
particularly appealing because it produces a real stigmatic image of a
homogeneous region with a reasonable range of refractive index, and we believe
that this lens can find important practical applications. Whether the imaging
devices discussed in this paper can provide super-resolution remains an open
question, but in our opinion the answer is positive.

\section*{Acknowledgements}
\label{}

This work was supported by the grants MSM0021622409 and MSM0021622419 of the
Czech Ministry of Education and by GA\v CR 202/08/H072.


\newcommand{\noi}{}
\newcommand{\beq}{\begin{equation}}
\newcommand{\eeq}{\end{equation}}
\newcommand{\bea}{\begin{eqnarray}}
\newcommand{\eea}{\end{eqnarray}}

\renewcommand{\e}[1]{{(\ref{#1})}}
\newcommand{\eq}[1]{{Eq.\ (\ref{#1})}}
\newcommand{\es}[2]{{(\ref{#1}) and (\ref{#2})}}
\newcommand{\eqs}[2]{{Eqs.\ (\ref{#1}) and (\ref{#2})}}
\newcommand{\Ref}[1]{{Ref.~\cite{#1}}}
\newcommand{\equi}[1]{\stackrel{{#1}}{=}}
\newcommand{\ie}{{${ i.e., \ }$}}
\newcommand{\eg}{{${ e.g., \ }$}}
\newcommand{\cf}{{cf.\ }}
\newcommand{\wrt}{{with respect to }}
\newcommand{\lhs}{{left-hand side }}
\newcommand{\rhs}{{right-hand side }}
\renewcommand{\~}{ \ }
\renewcommand{\=}{ \ = \ }
\newcommand{\for}{\mathrm{for}}
\newcommand{\Hf}{{\frac{1}{2}}}
\newcommand{\twobyone}[2]{\left(\begin{array}{c}{#1} \cr
                                {#2} \end{array} \right)}

\newcommand{\proofbox}{{\ \ $\ensuremath{\Box}$}}

\newtheorem{theorem}{Theorem}[section]
\newtheorem{assumption}[theorem]{Assumption}
\newtheorem{lemma}[theorem]{Lemma}

\section*{Appendix: No-Go Theorem for Spherically Symmetric Lenses}

In the following we show that the refractive index with the appealing
properties described in Sec.~\ref {tomandmartin} does not exist. This gives a
negative answer to the following question: is there a spherically symmetric
medium that focuses parallel rays in a spherical cavity via a shell-shaped lens
into a single point? If the answer were positive, this would constitute a new
interesting and practically important absolute instrument as well as a useful
building block for other optical devices \cite{Tyc2010}. Although the actual
answer is negative by itself, it still useful and likely to influence future
developments in the subject \cite{t11}.

Consider parallel rays located in an inner spherical region $r\!\leq\!r_{*}$
with a constant refractive index $n_{*}\!\equiv\!\frac{N_{*}}{r_{*}}\!>\!0$
that are focused by the medium to a point $P$ situated at some finite distance
$R\!>\!r_{*}$ from the centre $O$. We will try to determine an appropriate
spherically symmetric profile $N(r)\!=\!r n(r)$ in the intermediate shell
region \beq r_{*}\~\leq\~ r\~ \leq\~ R \label{rbounds}\eeq with inner and outer
radius $r_{*}>0$ and $R<\infty$, respectively. We will assume that all rays are
outgoing, \ie the radial position of a light ray $r(t)$ is monotonically
increasing as a function of time $t$, so no turning points are allowed. In
particular, the outer region $r\!>\!R$ will play no role, and we might as well
assume that the focal point P lies on the outer rim of the device. Suppose
that the parallel rays move horizontally to the left and are focused to the
focal point P with polar coordinates $(R,\pi)$.  As before, it is convenient to
introduce the logarithm $x\!\equiv\!\ln r$ of the radial coordinate
$r$. Similarly, we shall write $x_{*}\!\equiv\!\ln r_{*}$, $X\!\equiv\!\ln R$
etc.\ in an obvious notation.

Clearly $N_{*}$ is the upper bound for the angular momentum $L$ inside the
inner region $r\!\leq\!r_{*}$. However, we shall only require that the focusing
device actually works for rays with $L\!\in\![0,L_{m})$, where $L_{m}\!\in{}
  (0,N_{*}]$.  A ray with angular momentum $L$ emerges horizontally to the left
from the inner rim at a polar angle $\pi-\arcsin\frac{L}{N_{*}}$ and when it
arrives at P, the polar angle becomes $\pi$, so the corresponding change of
polar angle is $\arcsin\frac{L}{N_{*}}$. On the other hand, the same change can
be calculated by integrating Eq.~(\ref{dfidr}) from $r_*$ to $R$, so we arrive
at the following integral equation: \beq \mbox{$
\forall L\in[0,L_{m}):\~\~\frac{1}{L} \arcsin\frac{L}{N_{*}}
      \= \int_{r_{*}}^{R} \frac{\dd r}{r\sqrt{N^{2}(r)-L^{2}}} \~\equiv\~
      \int_{x_{*}}^{X}\frac{\dd x}{\sqrt{N^{2}(x)-L^{2}}}\~.$}
\label{sarborttycinteqgen}
\eeq
We next redefine (normalise) the quantities $N^{\prime}\!:=\!\frac{N}{N_{*}}$,
$L^{\prime}\!:=\!\frac{L}{N_{*}}$, $L^{\prime}_{m}\!:=\!\frac{L_{m}}{N_{*}}$, 
etc., by dividing with the constant $N_{*}$. Suppressing the primes from now
on, we arrive at the main integral equation \cite{Tyc2010}, 
\beq
\fbox{$
\forall L\in[0,L_{m}):\~\~
\frac{\arcsin(L)}{L} \= \int_{r_{*}}^{R} \frac{\dd r}{r\sqrt{N^{2}(r)-L^{2}}} 
\~\equiv\~ \int_{x_{*}}^{X}\frac{\dd x}{\sqrt{N^{2}(x)-L^{2}}} \~,$}
\label{sarborttycinteq}
\eeq
which serves as our starting point.
The solution function $N\!=\!N(r)$ is required to be bounded 
\beq
\exists N_{m}\,\,\forall r\in[r_{*},R]:\~\~
0<L_{m}\leq N(r) \leq N_{m} < \infty \label{nbounds}
\eeq
for physical reasons, and Lebesgue measurable \cite{rudin66} in order
for the integral \e{sarborttycinteq} to be defined. Since the integrand
\e{sarborttycinteq} is non-negative, the integral is always defined, 
although perhaps infinite. 
The goal is now to investigate the integral equation \e{sarborttycinteq}, and
show that it has no bounded solutions $N\!=\!N(r)$.

\noi
Note that the $k$'th moment $\int_{r_{*}}^{R}\frac{\dd r}{r} \frac{1}{N^{k}(r)}$ 
is well-defined and finite for any integer $k$ because of the bounds 
\es{rbounds}{nbounds}.

\begin{lemma}
[Reformulation in terms of odd moments]
The integral equation \e{sarborttycinteq} is equivalent to 
\beq
\fbox{$
\int_{r_{*}}^{R}\frac{\dd r}{r}\frac{1}{N^{2n+1}} \= \frac{1}{2n+1} 
\~\~\for \~\~n\in\mathbb{N}_{0}\equiv\{0,1,2,3,\ldots\}\~.$}
\label{moment}
\eeq
\end{lemma}

\noi
{\sc Proof:}\~\~The Taylor expansion of the $\arcsin(L)$ function at $L=0$ is
\beq
\mbox{$
\frac{\arcsin(L)}{L} \= \sum_{n=0}^{\infty}\frac{1}{2n+1}\twobyone{2n}{n}
\left(\frac{L}{2}\right)^{2n}\~.$}\label{lhsmoment}
\eeq
On the \rhs of \eq{sarborttycinteq}, a binomial expansion produces 
\beq
\begin{array}{rcl}
\int_{x_{*}}^{X}\frac{\dd x}{\sqrt{N^{2}-L^{2}}} 
&=& \int_{x_{*}}^{X}\frac{\dd x}{N}\left(1-(\frac{L}{N})^{2}\right)^{-\Hf}
\= \int_{x_{*}}^{X}\frac{\dd x}{N}\sum_{n=0}^{\infty}\twobyone{-\Hf}{n}
\left(-(\frac{L}{N})^{2}\right)^{n} \cr
&=& \int_{x_{*}}^{X}\frac{\dd x}{N}\sum_{n=0}^{\infty}\twobyone{2n}{n}
\left(\frac{L}{2N}\right)^{2n}
\=\sum_{n=0}^{\infty}\twobyone{2n}{n} 
\int_{x_{*}}^{X}\frac{\dd x}{N}\left(\frac{L}{2N}\right)^{2n}\~.
\end{array}
\label{rhsmoment}
\eeq
In the last equality of \eq{rhsmoment}, we were allowed to exchange 
integration and summation order, because each term is positive or zero 
(Tonelli's Theorem). In the third equality of \eq{rhsmoment}, we have used the
identity
\beq
\mbox{$
\twobyone{-\Hf}{n} 
\= \twobyone{2n}{n} \left(-\frac{1}{4}\right)^{n}\~.$}
\eeq
Comparing the Taylor-coefficients on the \lhs \e{lhsmoment} and \rhs 
\e{rhsmoment} of \eq{sarborttycinteq} yields the reformulation \e{moment}.
\proofbox

\begin{lemma}[Reformulation in terms of new $L$]
The integral equation \e{sarborttycinteq} is equivalent to 
\beq
\fbox{$
\forall L\in\mathbb{C}:\~
e^{L^{2}}
\=\int_{r_{*}}^{R}\frac{\dd r}{r N}
\left( 1+2(\frac{L}{N})^{2} \right)\exp\left[(\frac{L}{N})^{2}\right] 
\~\equiv\~
\left.\left( 1+\frac{\dd }{\dd \alpha}\right) \int_{r_{*}}^{R}\frac{\dd r}{r N}
\exp\left[(\frac{\alpha L}{N})^{2}\right] \right|_{\alpha=1}\~. $}
\label{newell}
\eeq
\end{lemma}

\noi
{\sc Proof:}\~\~The integrals in \eq{newell} are well-defined due to the
bounds \es{rbounds}{nbounds}. We calculate
\beq
\begin{array}{rcl}
&&L \int_{x_{*}}^{X}\frac{\dd x}{N}\exp\left[(\frac{L}{N})^{2}\right] 
\= L \int_{x_{*}}^{X}\frac{\dd x}{N}\sum_{n=0}^{\infty}\frac{1}{n!}
\left(\frac{L}{N}\right)^{2n}
\=\sum_{n=0}^{\infty}\frac{1}{n!} \int_{x_{*}}^{X}\! \dd x
\left(\frac{L}{N}\right)^{2n+1} \cr
&\equi{\e{moment}}&\sum_{n=0}^{\infty}\frac{1}{n!} \frac{L^{2n+1}}{2n+1}
\=\sum_{n=0}^{\infty}\frac{1}{n!} \int_{0}^{L}\! \dd \ell \~ \ell^{2n}
\= \int_{0}^{L}\! \dd \ell \sum_{n=0}^{\infty}\frac{1}{n!} \ell^{2n}
\= \int_{0}^{L}\! \dd \ell \~e^{\ell^{2}} \~.
\end{array}\label{newellint} 
\eeq
In the second equality of \eq{newellint}, we use Tonelli's and Fubini's 
Theorems \cite{rudin66} with $x$-independent majorant 
$\frac{|L|}{L_{m}}\exp\left[(\frac{|L|}{L_{m}})^{2}\right]$ to justify exchange
of integration and summation order. Tonelli's and Fubini's Theorems are also 
used in the fifth equality of \eq{newellint} with majorant $e^{|\ell|^{2}}$.
Now differentiate both sides of \eq{newellint} \wrt $L$ to obtain \eq{newell}.
Note that \eq{newellint} is trivially satisfied for $L=0$, so we do not
lose information when differentiating. Therefore we can also run the argument
backwards.
\proofbox

\noi
Thus we have three equivalent conditions, Eqs.\ \e{sarborttycinteq},
\es{moment}{newell}, that a solution $N\!=\!N(r)$ should satisfy. 

\begin{lemma}
If there exists a solution $N\!=\!N(r)$ (not necessarily monotonically 
increasing as a function of the radius $r$), then there also exists
a monotonically increasing solution $\acute{N}\!=\!\acute{N}(r)$.
\label{mono}
\end{lemma}

\noi
{\sc Sketched proof:}\~\~Lemma~\ref{mono} is basically the observation that
the integration variable $x\equiv\ln r$ only enters implicitly via the function
$N=N(x)$, and that the Lebesgue measure $\dd x$ is translation invariant.
\proofbox

\noi
{}From now on we can and we will make the following assumption~\ref{assump01}
without loss of generality.

\begin{assumption}
The solution $N\!=\!N(r)$ is a monotonically increasing function of $r$. 
\label{assump01}
\end{assumption}

\noi
At this point, we introduce a technical assumption~\ref{assump02} in order to
proceed.
 
\begin{assumption}
The inverse solution $r\!=\!r(N)$ exists and is differentiable with Lebesgue 
measurable derivative.
\label{assump02}
\end{assumption}

\noi
The Assumption~\ref{assump02} implies that one can define a Lebesgue 
measurable density 
\beq
\mbox{$
\rho(N)\~:=\~\frac{\dd \ln r(N)}{\dd \ln N} \~\geq\~ 0\~.$} \label{rhodef}
\eeq
Let us call the definition domain of the inverse solution $r=r(N)$ for 
$[N_{*},N_{m}]$. In other words, $r(N_{*})=r_{*}$ and $r(N_{m})=R$.

\begin{lemma}[Reformulation in terms of test functions]
Under the assumptions~\ref{assump01}-\ref{assump02},
the integral equation \e{sarborttycinteq} becomes equivalent to 
\beq
\fbox{$
\forall \eta\in C^{\infty}_{c}((0,\infty)):\~\~ 
\int_{N_{*}}^{N_{m}}\!\rho(N) \dd N \~\frac{\dd \eta(N)}{\dd N}\= -\eta(1)\~.$}
\label{testfctlemmaenn}
\eeq
\end{lemma}

\noi
{\sc Remark:}\~\~Here $C^{\infty}_{c}((0,\infty))$ denotes the set of infinitely
often differentiable functions $\eta$ defined on the open interval
$(0,\infty)$, and such that $\eta$ has compact support in $(0,\infty)$. 
Compact support means that the function $\eta$ is assumed to vanish
identically in whole neighbourhoods around of $N=0$ and $N=\infty$. It is
therefore natural to extend $\eta$ smoothly to the closed interval
$[0,\infty]$ by assigning to $\eta$ the values $\eta(0)=0=\eta(\infty)$ at the
end points $N=0$ and $N=\infty$.

\noi
{\sc Proof:}\~\~When we substitute the inverse solution $r=r(N)$,
equation \e{newell} becomes
\beq
\mbox{$e^{L^{2}}
\=\left.\left( 1+\frac{\dd }{\dd \alpha}\right) \int_{N_{*}}^{N_{m}}\!\rho(N) 
\frac{\dd N}{N^{2}}
\exp\left[(\frac{\alpha L}{N})^{2}\right]\right|_{\alpha=1}\~.$}
 \label{newellenn}
\eeq
Next perform the elementary substitution $\nu\equiv 1/N$ with limits 
$\nu_{m}\equiv 1/N_{m}$, and $\nu_{*}\equiv 1/N_{*}$. Furthermore, multiply 
both sides with $e^{-(L\mu)^{2}}$, where $\mu>0$ is a positive parameter. Then 
\beq
\mbox{$e^{L^{2}(1-\mu^{2})}
\=\left.\left( 1+\frac{\dd }{\dd \alpha}\right)
\int_{\nu_{m}}^{\nu_{*}}\!\rho(\nu) 
\dd \nu\~e^{L^{2}((\alpha\nu)^{2}-\mu^{2})} \right|_{\alpha=1}\~.$}
 \label{newellnu}
\eeq
Recall that the Dirac delta distribution $\delta(x)$ has the Fourier integral 
representation $\delta(x)=\int_{-\infty}^{\infty}\frac{\dd p}{2\pi}e^{\ii px}$. 
By integrating $L^{2}\=\ii p$ along the imaginary axis in \eq{newellnu}, one
gets
\beq
\mbox{$\delta(1-\mu^{2})
\=\left.\left( 1+\frac{\dd }{\dd \alpha}\right)
\int_{\nu_{m}}^{\nu_{*}}\!\rho(\nu) 
\dd \nu\~\delta((\alpha\nu)^{2}-\mu^{2}) \right|_{\alpha=1}\~.$}
 \label{newelldelta2}  
\eeq
By using the Jacobian formula for the Dirac delta distribution
\beq 
\mbox{$\delta(f(x))
\=\sum_{\footnotesize \begin{array}{c} x_{0} \cr f(x_{0})=0\end{array}}
\frac{1}{|f^{\prime}(x_{0})|}\~\delta(x-x_{0})\~,$}
\eeq
and multiplying both sides with $2$, one gets 
\beq
\mbox{$\delta(\mu-1)
\=\left.\left( 1+\frac{\dd }{\dd \alpha}\right) 
\int_{\nu_{m}}^{\nu_{*}}\!\rho(\nu) 
\dd \nu\~\frac{1}{\mu}\delta(\alpha\nu-\mu) \right|_{\alpha=1}\~,$}
 \label{newelldelta1}
\eeq
where we have assumed that $\mu>0$ is positive.
Thus for a test function $\eta\in C^{\infty}_{c}((0,\infty))$, one calculates
\beq
\begin{array}{rcl}
\eta(1)&=& \int_{0}^{\infty} \! \dd \mu \~\eta(\mu) \delta(\mu-1) 
\~\equi{\e{newelldelta1}}\~
\left.\left( 1+\frac{\dd }{\dd \alpha}\right) \int_{\nu_{m}}^{\nu_{*}}\!\rho(\nu) 
\dd \nu \int_{0}^{\infty} \! \dd \mu \~
\frac{\eta(\mu)}{\mu}\delta(\alpha\nu-\mu)
\right|_{\alpha=1} \cr
&=& \left.\left( 1+\frac{\dd }{\dd \alpha}\right) 
\int_{\nu_{m}}^{\nu_{*}}\!\rho(\nu) 
\dd \nu  \~\frac{\eta(\alpha\nu )}{\alpha\nu}
\right|_{\alpha=1}
\= \left.\int_{\nu_{m}}^{\nu_{*}}\!\rho(\nu)\dd \nu 
\left[\frac{\eta(\alpha\nu )}{\alpha\nu}
- \frac{\eta(\alpha\nu )}{\alpha^{2}\nu}+
\frac{\eta^{\prime}(\alpha\nu)}{\alpha} \right] \right|_{\alpha=1} \cr
&=&\int_{\nu_{m}}^{\nu_{*}}\!\rho(\nu)\dd \nu \frac{\dd \eta(\nu)}{\dd \nu}\~.
\end{array}
\label{testfctlemmanu}
\eeq
Now translate \e{testfctlemmanu} back to the $N\equiv 1/\nu$ variable to
obtain \eq{testfctlemmaenn}.
\proofbox

\begin{lemma}
\eq{testfctlemmaenn} has no solutions for $\rho$ that respects the bounds
\e{nbounds} on $N$.
\end{lemma}

\noi
{\sc Proof:}\~\~The Fundamental Lemma of calculus of variation (in the 
strengthen version of du Bois-Reymond) \cite{hoerm90} shows that $\rho$ must 
be a constant up to contributions that vanish almost everywhere. 
(In particular, we stress that it is not enough for $\rho$ to be only piecewise
constant.) Thus one may pull the density $\rho$ outside of the integral 
\e{testfctlemmaenn}, and integrate to get
\beq
\forall\eta\in C^{\infty}_{c}((0,\infty)):
\~\~\rho \left( \eta(N_{m})-\eta(N_{*})\right)\= -\eta(1) \~.
\label{testfctlemmaennint}
\eeq
Collapsing limits $N_{*}\!=\!N_{m}$ are clearly not a solution. Assuming 
$N_{*}\!<\! N_{m}$, equation \e{testfctlemmaennint} has two solutions, 
$(\rho\!=\!1, N_{*}\!=\!1, N_{m}\!=\!\infty)$, or 
$(\rho\!=\!-1, N_{*}\!=\!0, N_{m}\!=\!1)$.
However, none of these two solutions respect the bounds \e{nbounds} 
on $N$, and the latter is actually monotonically decreasing.
\proofbox


\begin{thebibliography}{20}
\bibitem{Pendry2000} Pendry J B 2000 Phys. Rev. Lett. \textbf{85} 3966
\bibitem{Soukoulis2007} Soukoulis C M, Linden S and Wegener M 2007 Science \textbf{315} 47
\bibitem{Fang2005-neg_n-silver_imaging} Fang N \textit{et al} 2005 Science \textbf{308} 534
\bibitem{Stockman2007} Stockman M I 2007 Phys. Rev. Lett. \textbf{98} 177404
\bibitem{Ulf2009-fisheye} Leonhardt U 2009 New J. Phys. \textbf{11} 093040
\bibitem{Ulf2010-fisheye} Leonhardt U and Philbin T G 2010 Phys. Rev. A \textbf{81} 011804(R)
\bibitem{Ma2011} Ma Y G \textit{et al} 2011 New J. Phys. \textbf{13} 033016 
\bibitem{MFE} Maxwell J C 1854 Camb. Dublin Math. J. \textbf{8} 188
\bibitem{BornWolf} Born M and Wolf E 2006 \textit{Principles of optics} (Cambridge: Cambridge University Press)
\bibitem{Hendi2006} Hendi A, Henn J and Leonhardt U 2006 Phys. Rev. Lett \textbf{97} 073902
\bibitem{Ulf-Thomas-book} Leonhardt U and Philbin T 2010 \textit{Geometry and Light: The Science of Invisibility} (Dover: Mineola)
\bibitem{Firsov1953} Firsov O B 1953 Zh. Eksp. Teor. Fiz. \textbf{24}, 279
\bibitem{Luneburg1964} Luneburg R K 1964 \textit{Mathematical Theory of Optics} (Berkeley: University of California Press)
\bibitem{Ost97} Ostrovsky V N 1997 Phys. Rev. A \textbf{56} 526
\bibitem{Landau} Landau L D and Lifshitz E M 1976 \textit{Mechanics} (Oxford: Butterworth-Heinemann)
\bibitem{Demkov1971} Demkov Y N and Ostrovsky V N 1971  Sov. Phys.-JETP \textbf{33} 1083
\bibitem{Eaton1952} Eaton J E 1952 Trans. IRE Ant. Prop. \textbf{4} 66
\bibitem{Minano2006} Mi\~nano J C 2006 Opt. Express \textbf{14} 9627
\bibitem{Tyc2010} Tyc T and \v Sarbort M 2010 {arXiv:1010.3178}

\bibitem{t11} Tyc~T 2011 arXiv:1103.3406

\bibitem{rudin66}
Rudin~W 1966 {\em Real and complex analysis} (New York: McGraw-Hill) 

\bibitem{hoerm90}
H\"ormander~L 1990 
{\em The Analysis of Linear Partial Differential Operators I, 
(Distribution theory and Fourier Analysis)} (Springer-Verlag)


\end{thebibliography}
\end{document}